\documentclass[runningheads]{llncs}

\usepackage[utf8]{inputenc}
\usepackage[usenames,dvipsnames,table]{xcolor}
\usepackage{orcidlink}

\usepackage{breakurl}
\usepackage{eucal}

\usepackage{centernot}
\usepackage{mdframed}

\usepackage{paralist}
\usepackage{mathtools}
\usepackage{url}
\usepackage{extarrows}

\usepackage{wasysym} 
\usepackage{newunicodechar}
\newunicodechar{€}{\euro}
\usepackage{etoolbox}
\usepackage{xspace} 

\usepackage{stmaryrd} 

\usepackage[inline,shortlabels]{enumitem} 
\newlist{inlinelist}{enumerate*}{1}
\setlist*[inlinelist,1]{%
  label=(\roman*),
}

\usepackage{xifthen}  
\usepackage{ifthen}

\usepackage{nicefrac}

\usepackage{color}

\usepackage{tikz} 
\usepackage{float}
\usepackage{hhline}

\usepackage{stackengine} 

\usepackage{pifont}

\usepackage{hyperref}
\usepackage{cleveref}

\usepackage[final,nomargin,inline,index]{fixme} 
\fxusetheme{color}

\definecolor{Black}{HTML}{000000}
\definecolor{Gray}{HTML}{808080}
\definecolor{Magenta}{HTML}{FF00FF}
\definecolor{RubineRed}{HTML}{ED017D}
\definecolor{ForestGreen}{HTML}{028A0F}
\definecolor{MidnightBlue}{HTML}{006795}
\definecolor{Plum}{HTML}{92268F}

\FXRegisterAuthor{bart}{anbart}{\color{magenta} {\underline{bart}}}
\FXRegisterAuthor{enrico}{anel}{\color{red} {\underline{enrico}}}
\FXRegisterAuthor{livio}{anlivio}{\color{blue} {\underline{livio}}}

\hypersetup{
  breaklinks   = true,
  colorlinks   = true, 
  urlcolor     = blue, 
  linkcolor    = blue, 
  citecolor    = red   
}
\hypersetup{final} 

\usepackage{caption}
\DeclareCaptionType{listing}[Listing][List of Listings] 
\crefname{listing}{Listing}{Listings}  

\usepackage[most]{tcolorbox}

\tcbset{myoneside/.style={
colback=LightGrey,
colframe=gray,
top=0pt,
left=2pt,
right=2pt,
bottom=-5pt,
boxsep=1pt,
halign=flush center,
nobeforeafter,
before skip=-5pt,
after skip=0pt,
boxrule=0pt
}}

\tcbset{mytwosides/.style={
colback=LightGrey,
colframe=gray,
top=0pt,
left=2pt,
right=2pt,
bottom=-5pt,
boxsep=1pt,
sidebyside,
sidebyside align=top,
sidebyside gap=8pt,
halign=flush center,
nobeforeafter,
before skip=-5pt,
after skip=0pt,
boxrule=1pt
}}

\usepackage{listings}
\definecolor{listingBG}{HTML}{FFFFCB}%
\definecolor{listingFrame}{HTML}{BBBB98}%
\definecolor{listingLineno}{rgb}{0.5,0.5,1.0}%

\definecolor{LightGrey}{rgb}{0.975,0.975,0.975}

\lstdefinelanguage{solidity}{
	commentstyle=\color{Gray},
	morecomment=[l]{//},
	morecomment=[s]{/*}{*/},
	classoffset=0,
        escapechar=\$,
	morekeywords={anonymous, assembly, balance, break, call, callcode, case, catch, class, constant, continue, constructor, contract, debugger, default, delegatecall, delete, do, else, emit, event, experimental, export, external, false, finally, for, function, gas, if, implements, import, in, indexed, instanceof, interface, internal, is, length, library, log0, log1, log2, log3, log4, memory, modifier, new, payable, pragma, private, protected, public, pure, push, return, returns, revert, rule, selfdestruct, send, solidity, storage, struct, suicide, super, switch, then, this, throw, transfer, true, try, typeof, using, value, view, while, with, addmod, ecrecover, keccak256, mulmod, ripemd160, sha256, sha3},
	keywordstyle=\color{NavyBlue}\bfseries,
	classoffset=1,
	morekeywords={address, bool, byte, bytes, bytes1, bytes2, bytes3, bytes4, bytes5, bytes6, bytes7, bytes8, bytes9, bytes10, bytes11, bytes12, bytes13, bytes14, bytes15, bytes16, bytes17, bytes18, bytes19, bytes20, bytes21, bytes22, bytes23, bytes24, bytes25, bytes26, bytes27, bytes28, bytes29, bytes30, bytes31, bytes32, enum, int, int8, int16, int24, int32, int40, int48, int56, int64, int72, int80, int88, int96, int104, int112, int120, int128, int136, int144, int152, int160, int168, int176, int184, int192, int200, int208, int216, int224, int232, int240, int248, int256, mapping, string, uint, uint8, uint16, uint24, uint32, uint40, uint48, uint56, uint64, uint72, uint80, uint88, uint96, uint104, uint112, uint120, uint128, uint136, uint144, uint152, uint160, uint168, uint176, uint184, uint192, uint200, uint208, uint216, uint224, uint232, uint240, uint248, uint256, var, void, ether, finney, szabo, wei, days, hours, minutes, seconds, weeks, years},
	keywordstyle=\color{blue},
	classoffset=2,
	morekeywords={block, blockhash, coinbase, difficulty, gaslimit, number, timestamp, msg, data, gas, sender, sig, value, now, tx, gasprice, origin},
	keywordstyle=\color{Plum}\bfseries,
 	classoffset=3,
 	morekeywords={assert,require},
	keywordstyle=\color{red}\bfseries,
}

\lstdefinelanguage{cvl}{
	commentstyle=\color{Gray},
	morecomment=[l]{//},
	morecomment=[s]{/*}{*/},
	classoffset=0,
        escapechar=\$,
	morekeywords={anonymous, assembly, balance, break, call, callcode, case, catch, class, constant, continue, constructor, contract, debugger, default, delegatecall, delete, do, else, emit, event, experimental, export, external, false, finally, for, function, gas, if, implements, import, in, indexed, instanceof, interface, internal, is, length, library, log0, log1, log2, log3, log4, memory, modifier, new, payable, pragma, private, protected, public, pure, push, return, returns, revert,  selfdestruct, send, solidity, storage, struct, suicide, super, switch, then, this, throw, transfer, true, try, typeof, using, value, view, while, with, addmod, ecrecover, keccak256, mulmod, ripemd160, sha256, sha3},
	keywordstyle=\color{NavyBlue}\bfseries,
	classoffset=1,
	morekeywords={address, bool, byte, bytes, bytes1, bytes2, bytes3, bytes4, bytes5, bytes6, bytes7, bytes8, bytes9, bytes10, bytes11, bytes12, bytes13, bytes14, bytes15, bytes16, bytes17, bytes18, bytes19, bytes20, bytes21, bytes22, bytes23, bytes24, bytes25, bytes26, bytes27, bytes28, bytes29, bytes30, bytes31, bytes32, enum, int, int8, int16, int24, int32, int40, int48, int56, int64, int72, int80, int88, int96, int104, int112, int120, int128, int136, int144, int152, int160, int168, int176, int184, int192, int200, int208, int216, int224, int232, int240, int248, int256, mapping, string, uint, uint8, uint16, uint24, uint32, uint40, uint48, uint56, uint64, uint72, uint80, uint88, uint96, uint104, uint112, uint120, uint128, uint136, uint144, uint152, uint160, uint168, uint176, uint184, uint192, uint200, uint208, uint216, uint224, uint232, uint240, uint248, uint256, var, void, ether, finney, szabo, wei, days, hours, minutes, seconds, weeks, years},
	keywordstyle=\color{blue},
	classoffset=2,
	morekeywords={block, blockhash, coinbase, difficulty, gaslimit, number, timestamp, msg, data, gas, sender, sig, value, now, tx, gasprice, origin},
	keywordstyle=\color{Plum}\bfseries,
 	classoffset=3,
 	morekeywords={invariant,rule,assert,satisfy,require},
	keywordstyle=\color{red}\bfseries,
}
\usepackage{tabularx}
\usepackage{makecell}
\usepackage{fp}  
\usepackage{multirow}

\newcommand{\change}[1]{#1}

\newcommand{\ifempty}[3]{%
  \ifthenelse{\isempty{#1}}{#2}{#3}%
}

\newcommand{\ifdots}[3]{%
  \ifthenelse{\equal{#1}{...}}{#2}{#3}%
}

\newcommand{\hidden}[1]{}

\makeatletter
\newcommand*{\itemequation}[3][]{%
  \item
  \begingroup
    \refstepcounter{equation}%
    \ifx\\#1\\%
    \else  
      \label{#1}%
    \fi
    \sbox0{#2}%
    \sbox2{$\displaystyle#3\m@th$}%
    \sbox4{\@eqnnum}%
    \dimen@=.5\dimexpr\linewidth-\wd2\relax
    \ifcase
        \ifdim\wd0>\dimen@
          \z@
        \else
          \ifdim\wd4>\dimen@
            \z@
          \else 
            \@ne
          \fi 
        \fi
      \@latex@warning{Equation is too large}%
    \fi
    \noindent   
    \rlap{\copy0}%
    \rlap{\hbox to \linewidth{\hfill\copy2\hfill}}%
    \hbox to \linewidth{\hfill\copy4}%
    \hspace{0pt}
  \endgroup
  \ignorespaces 
}
\makeatother

\newcommand{\Real}[1]{\mathrm{Real}}

\newcommand{\codefont}{\fontsize{9}{9}\selectfont}
\newcommand{\code}[1]{{\tt\codefont{#1}}}

\def\etc{etc.\@\xspace}

\newcommand{\eg}{e.g.\@\xspace}
\newcommand{\ie}{i.e.\@\xspace}
\newcommand{\wrt}{w.r.t.\@\xspace}

  {}

\DeclareMathAlphabet{\mathbfsf}{\encodingdefault}{\sfdefault}{bx}{n}

\newcommand{\waldistrarrow}[1]{\approx_{\$}}

\definecolor{LightGrey}{rgb}{0.95,0.95,0.95}
\definecolor{keyword}{HTML}{7F0055}

\newlength\replength
\newcommand\repfrac{.1}

\newcommand\rulewidth{.6pt}
\newcommand\tdashfill[1][\repfrac]{\cleaders\hbox to \replength{%
  \smash{\rule[\arraystretch\ht\strutbox]{\repfrac\replength}{\rulewidth}}}\hfill}

\newcommand\tdotfill[1][\repfrac]{\cleaders\hbox to \replength{%
  \smash{\raisebox{\arraystretch\dimexpr\ht\strutbox-.1ex\relax}{.}}}\hfill}

\newcommand{\contrAdvC}[2]{\mathcal{C}} 

\newcommand{\gptC}{{GPT-5}\xspace}
\newcommand{\gptQ}{GPT-4\xspace}
\newcommand{\solcmc}{SolCMC\xspace}
\newcommand{\certora}{Certora\xspace}

\newcommand{\propname}[1]{``\emph{#1}''}

\newcommand{\False}{\textit{False}\xspace}
\newcommand{\True}{\textit{True}\xspace}
\newcommand{\Unknown}{\textit{Unknown}\xspace}

\newcommand{\dataset}{\ensuremath{\mathrm{DS}}\xspace}
\newcommand{\datasetCVL}{\ensuremath{\mathrm{\dataset_{|CVL}}}\xspace}
\newcommand{\datasetNoCVL}{\ensuremath{\mathrm{\dataset_{|\neg CVL}}}\xspace}

\newcommand{\datasetSolCMC}{\ensuremath{\mathrm{\dataset_{|SolCMC}}}\xspace}
\newcommand{\datasetNoSolCMC}{\ensuremath{\mathrm{\dataset_{|\neg SolCMC}}}\xspace}

\newcommand{\colorgptC}{cyan!20}
\newcommand{\colorgptQ}{gray!20}
\newcommand{\colorSol}{green!20}
\newcommand{\colorCert}{brown!20}

\newcommand{\rowgptC}{\rowcolor{\colorgptC}}
\newcommand{\rowgptQ}{\rowcolor{\colorgptQ}}
\newcommand{\rowSol}{\rowcolor{\colorSol}}
\newcommand{\rowCert}{\rowcolor{\colorCert}}

\newcolumntype{\columngptQ}{>{\columncolor{\colorgptQ}}c}
\newcolumntype{\columngptC}{>{\columncolor{\colorgptC}}c}
\newcolumntype{\columnSol}{>{\columncolor{\colorSol}}c}
\newcolumntype{\columnCert}{>{\columncolor{\colorCert}}c}

\usepackage{transparent}
\newcolumntype{\columngptQExpr}{>{\transparent{0.7}}>{\em}>{\columncolor{\colorgptQ}}c}
\newcolumntype{\columngptCExpr}{>{\transparent{0.7}}>{\em}>{\columncolor{\colorgptC}}c}

\newcommand{\exprcov}[2]{{#1}\quad ({#2})}

\usepackage[colorinlistoftodos,prependcaption,textsize=tiny]{todonotes}

\newtoggle{anonymous}

\begin{document}
%

\title{LLMs as verification oracles for Solidity}
%

\iftoggle{anonymous}{%
\author{Anonymous authors}
}
{%
\author{Massimo Bartoletti
 \orcidlink{0000-0003-3796-9774} 
\and 
Enrico Lipparini
 \orcidlink{0009-0009-0428-4403} 
\and 
Livio Pompianu
\orcidlink{0000-0003-3745-4324}
}
\institute{University of Cagliari, Italy}
}
%
%
\maketitle              

\newcommand{\Nusecases}{5}

\newcommand{\NvBank}{17}
\newcommand{\NpBank}{27}
\newcommand{\NlBank}{28}

\newcommand{\NvVault}{11}
\newcommand{\NpVault}{38}
\newcommand{\NlVault}{55}

\newcommand{\NvPriceBet}{16}
\newcommand{\NpPriceBet}{42}
\newcommand{\NlPriceBet}{73}

\newcommand{\NvPaymentSplitter}{10}
\newcommand{\NpPaymentSplitter}{17}
\newcommand{\NlPaymentSplitter}{145}

\newcommand{\NvLP}{9}
\newcommand{\NpLP}{35}
\newcommand{\NlLP}{303}

\newcommand{\Nversions}{\the\numexpr \NvBank + \NvVault + \NvPriceBet + \NvPaymentSplitter + \NvLP \relax}
\newcommand{\Nproperties}{\the\numexpr \NpBank + \NpVault + \NpPriceBet + \NpPaymentSplitter + \NpLP \relax}

\newcommand{\NvertasksBank}{\the\numexpr \NvBank*\NpBank\relax}
\newcommand{\NvertasksVault}{\the\numexpr \NvVault*\NpVault\relax}
\newcommand{\NvertasksPrice}{\the\numexpr \NvPriceBet*\NpPriceBet\relax}
\newcommand{\NvertasksSplit}{\the\numexpr \NvPaymentSplitter*\NpPaymentSplitter\relax}
\newcommand{\NvertasksLend}{\the\numexpr \NvLP*\NpLP\relax}

\newcommand{\Nvertasks}{\the\numexpr \NvertasksBank + \NvertasksVault + \NvertasksPrice + \NvertasksSplit + \NvertasksLend \relax}

\newcommand{\NvertasksEffBank}{121}
\newcommand{\NvertasksEffVault}{106}
\newcommand{\NvertasksEffPrice}{158}
\newcommand{\NvertasksEffSplit}{160}
\newcommand{\NvertasksEffLend}{122} 

\newcommand{\NvertasksEff}{\the\numexpr \NvertasksEffBank + \NvertasksEffVault + \NvertasksEffPrice + \NvertasksEffSplit + \NvertasksEffLend \relax}


\newcommand{\anBankFour}{121}
\newcommand{\unBankFour}{0}
\newcommand{\TPBankFour}{39}
\newcommand{\TNBankFour}{47}
\newcommand{\FPBankFour}{21}
\newcommand{\FNBankFour}{14}
\newcommand{\acBankFour}{71\%}
\newcommand{\prBankFour}{65\%}
\newcommand{\reBankFour}{74\%}
\newcommand{\spBankFour}{69\%}
\newcommand{\fsBankFour}{69\%}

\newcommand{\anVaultFour}{106}
\newcommand{\unVaultFour}{4}
\newcommand{\TPVaultFour}{37}
\newcommand{\TNVaultFour}{30}
\newcommand{\FPVaultFour}{25}
\newcommand{\FNVaultFour}{10}
\newcommand{\acVaultFour}{66\%}
\newcommand{\prVaultFour}{60\%}
\newcommand{\reVaultFour}{79\%}
\newcommand{\spVaultFour}{55\%}
\newcommand{\fsVaultFour}{68\%}

\newcommand{\anBetFour}{158}
\newcommand{\unBetFour}{4}
\newcommand{\TPBetFour}{39}
\newcommand{\TNBetFour}{55}
\newcommand{\FPBetFour}{34}
\newcommand{\FNBetFour}{26}
\newcommand{\acBetFour}{61\%}
\newcommand{\prBetFour}{53\%}
\newcommand{\reBetFour}{60\%}
\newcommand{\spBetFour}{62\%}
\newcommand{\fsBetFour}{57\%}

\newcommand{\anSplitFour}{160}
\newcommand{\unSplitFour}{2}
\newcommand{\TPSplitFour}{44}
\newcommand{\TNSplitFour}{43}
\newcommand{\FPSplitFour}{40}
\newcommand{\FNSplitFour}{31}
\newcommand{\acSplitFour}{55\%}
\newcommand{\prSplitFour}{52\%}
\newcommand{\reSplitFour}{59\%}
\newcommand{\spSplitFour}{52\%}
\newcommand{\fsSplitFour}{55\%}

\newcommand{\anLendFour}{122}
\newcommand{\unLendFour}{1}
\newcommand{\TPLendFour}{52}
\newcommand{\TNLendFour}{28}
\newcommand{\FPLendFour}{11}
\newcommand{\FNLendFour}{30}
\newcommand{\acLendFour}{66\%}
\newcommand{\prLendFour}{83\%}
\newcommand{\reLendFour}{63\%}
\newcommand{\spLendFour}{72\%}
\newcommand{\fsLendFour}{72\%}

\newcommand{\anOverallFour}{667}
\newcommand{\unOverallFour}{11}
\newcommand{\TPOverallFour}{211}
\newcommand{\TNOverallFour}{203}
\newcommand{\FPOverallFour}{131}
\newcommand{\FNOverallFour}{111}
\newcommand{\acOverallFour}{63\%}
\newcommand{\prOverallFour}{62\%}
\newcommand{\reOverallFour}{66\%}
\newcommand{\spOverallFour}{61\%}
\newcommand{\fsOverallFour}{64\%}

\newcommand{\anBankFive}{121}
\newcommand{\unBankFive}{0}
\newcommand{\TPBankFive}{50}
\newcommand{\TNBankFive}{65}
\newcommand{\FPBankFive}{3}
\newcommand{\FNBankFive}{3}
\newcommand{\acBankFive}{95\%}
\newcommand{\prBankFive}{94\%}
\newcommand{\reBankFive}{94\%}
\newcommand{\spBankFive}{96\%}
\newcommand{\fsBankFive}{94\%}

\newcommand{\anVaultFive}{106}
\newcommand{\unVaultFive}{0}
\newcommand{\TPVaultFive}{45}
\newcommand{\TNVaultFive}{53}
\newcommand{\FPVaultFive}{3}
\newcommand{\FNVaultFive}{5}
\newcommand{\acVaultFive}{92\%}
\newcommand{\prVaultFive}{94\%}
\newcommand{\reVaultFive}{90\%}
\newcommand{\spVaultFive}{95\%}
\newcommand{\fsVaultFive}{92\%}

\newcommand{\anBetFive}{158}
\newcommand{\unBetFive}{0}
\newcommand{\TPBetFive}{60}
\newcommand{\TNBetFive}{84}
\newcommand{\FPBetFive}{8}
\newcommand{\FNBetFive}{6}
\newcommand{\acBetFive}{91\%}
\newcommand{\prBetFive}{88\%}
\newcommand{\reBetFive}{91\%}
\newcommand{\spBetFive}{91\%}
\newcommand{\fsBetFive}{90\%}

\newcommand{\anSplitFive}{160}
\newcommand{\unSplitFive}{0}
\newcommand{\TPSplitFive}{69}
\newcommand{\TNSplitFive}{73}
\newcommand{\FPSplitFive}{14}
\newcommand{\FNSplitFive}{4}
\newcommand{\acSplitFive}{89\%}
\newcommand{\prSplitFive}{83\%}
\newcommand{\reSplitFive}{95\%}
\newcommand{\spSplitFive}{84\%}
\newcommand{\fsSplitFive}{88\%}

\newcommand{\anLendFive}{122}
\newcommand{\unLendFive}{0}
\newcommand{\TPLendFive}{79}
\newcommand{\TNLendFive}{37}
\newcommand{\FPLendFive}{3}
\newcommand{\FNLendFive}{3}
\newcommand{\acLendFive}{95\%}
\newcommand{\prLendFive}{96\%}
\newcommand{\reLendFive}{96\%}
\newcommand{\spLendFive}{93\%}
\newcommand{\fsLendFive}{96\%}

\newcommand{\anOverallFive}{667}
\newcommand{\unOverallFive}{0}
\newcommand{\TPOverallFive}{303}
\newcommand{\TNOverallFive}{312}
\newcommand{\FPOverallFive}{31}
\newcommand{\FNOverallFive}{21}
\newcommand{\acOverallFive}{92\%}
\newcommand{\prOverallFive}{91\%}
\newcommand{\reOverallFive}{94\%}
\newcommand{\spOverallFive}{91\%}
\newcommand{\fsOverallFive}{92\%}


\newcommand{\anaTrueBankFour}{86}
\newcommand{\corTrueBankFour}{91\%}
\newcommand{\comTrueBankFour}{99\%}
\newcommand{\cohTrueBankFour}{98\%}
\newcommand{\anaFalseBankFour}{35}
\newcommand{\corFalseBankFour}{6\%}
\newcommand{\comFalseBankFour}{37\%}
\newcommand{\cohFalseBankFour}{92\%}

\newcommand{\anaTrueOverallFour}{121} 

\newcommand{\corTrueBankFourgr}{78}
\newcommand{\comTrueBankFourgr}{85}
\newcommand{\cohTrueBankFourgr}{84}
\newcommand{\corFalseBankFourgr}{2}
\newcommand{\comFalseBankFourgr}{13}
\newcommand{\cohFalseBankFourgr}{32}

\newcommand{\anaTrueBankFive}{115}
\newcommand{\corTrueBankFive}{100\%}
\newcommand{\comTrueBankFive}{91\%}
\newcommand{\cohTrueBankFive}{99\%}
\newcommand{\anaFalseBankFive}{6}
\newcommand{\corFalseBankFive}{17\%}
\newcommand{\comFalseBankFive}{67\%}
\newcommand{\cohFalseBankFive}{83\%}

\newcommand{\corTrueBankFivegr}{115}
\newcommand{\comTrueBankFivegr}{105}
\newcommand{\cohTrueBankFivegr}{114}
\newcommand{\corFalseBankFivegr}{1}
\newcommand{\comFalseBankFivegr}{4}
\newcommand{\cohFalseBankFivegr}{5}

\newcommand{\anaTrueVaultFive}{54}
\newcommand{\corTrueVaultFive}{100\%}
\newcommand{\comTrueVaultFive}{91\%}
\newcommand{\cohTrueVaultFive}{98\%}
\newcommand{\anaFalseVaultFive}{8}
\newcommand{\corFalseVaultFive}{63\%}
\newcommand{\comFalseVaultFive}{88\%}
\newcommand{\cohFalseVaultFive}{50\%}

\newcommand{\corTrueVaultFivegr}{54}
\newcommand{\comTrueVaultFivegr}{49}
\newcommand{\cohTrueVaultFivegr}{53}
\newcommand{\corFalseVaultFivegr}{5}
\newcommand{\comFalseVaultFivegr}{7}
\newcommand{\cohFalseVaultFivegr}{4}

\newcommand{\anaTrueBetFive}{74}
\newcommand{\corTrueBetFive}{99\%}
\newcommand{\comTrueBetFive}{89\%}
\newcommand{\cohTrueBetFive}{100\%}
\newcommand{\anaFalseBetFive}{10}
\newcommand{\corFalseBetFive}{70\%}
\newcommand{\comFalseBetFive}{50\%}
\newcommand{\cohFalseBetFive}{80\%}

\newcommand{\corTrueBetFivegr}{73}
\newcommand{\comTrueBetFivegr}{66}
\newcommand{\cohTrueBetFivegr}{74}
\newcommand{\corFalseBetFivegr}{7}
\newcommand{\comFalseBetFivegr}{5}
\newcommand{\cohFalseBetFivegr}{8}

\newcommand{\anaTrueSplitFive}{88}
\newcommand{\anaFalseSplitFive}{17}

\newcommand{\corTrueSplitFivegr}{85}
\newcommand{\comTrueSplitFivegr}{83}
\newcommand{\cohTrueSplitFivegr}{84}
\newcommand{\corFalseSplitFivegr}{3}
\newcommand{\comFalseSplitFivegr}{15}
\newcommand{\cohFalseSplitFivegr}{14}

\newcommand{\anaTrueLendFive}{56}
\newcommand{\anaFalseLendFive}{5}

\newcommand{\corTrueLendFivegr}{52}
\newcommand{\comTrueLendFivegr}{48}
\newcommand{\cohTrueLendFivegr}{56}
\newcommand{\corFalseLendFivegr}{3}
\newcommand{\comFalseLendFivegr}{3}
\newcommand{\cohFalseLendFivegr}{2}

\newcommand{\anaTrueOverallFive}{304} 


\newcommand{\exprSolBank}{62.81}
\newcommand{\covSolBank}{61.98}
\newcommand{\exprCertBank}{89.26}
\newcommand{\covCertBank}{89.26}
\newcommand{\covGptBank}{100.00}

\newcommand{\exprSolVault}{26.42}
\newcommand{\covSolVault}{26.42}
\newcommand{\exprCertVault}{54.72}
\newcommand{\covCertVault}{54.72}
\newcommand{\covGptVault}{100.00}

\newcommand{\exprSolPrice}{47.47}
\newcommand{\covSolPrice}{46.84}
\newcommand{\exprCertPrice}{56.33}
\newcommand{\covCertPrice}{56.33}
\newcommand{\covGptPrice}{100.00}

\newcommand{\exprSolSplit}{43.75}
\newcommand{\covSolSplit}{18.13}
\newcommand{\exprCertSplit}{87.50}
\newcommand{\covCertSplit}{87.50}
\newcommand{\covGptSplit}{100.00}

\newcommand{\exprSolLend}{4.10}
\newcommand{\covSolLend}{0} 
\newcommand{\exprCertLend}{81.15}
\newcommand{\covCertLend}{77.05}
\newcommand{\covGptLend}{100.00}

\newcommand{\exprSolTot}{%
	\FPeval\tmp{
		(
		\exprSolBank*\NvertasksEffBank + 
		\exprSolVault*\NvertasksEffVault  + \exprSolPrice*\NvertasksEffPrice +
		\exprSolSplit*\NvertasksEffSplit + \exprSolLend*\NvertasksEffLend  	
		) / \NvertasksEff
	}%
	\FPround\tmp\tmp{2}\tmp
}

\newcommand{\covSolTot}{%
	\FPeval\tmp{
		(
		\covSolBank*\NvertasksEffBank + 
		\covSolVault*\NvertasksEffVault  + \covSolPrice*\NvertasksEffPrice +
		\covSolSplit*\NvertasksEffSplit + \covSolLend*\NvertasksEffLend  	
		) / \NvertasksEff
	}%
	\FPround\tmp\tmp{2}\tmp
}

\newcommand{\exprCertTot}{%
	\FPeval\tmp{
		(
		\exprCertBank*\NvertasksEffBank + 
		\exprCertVault*\NvertasksEffVault  + \exprCertPrice*\NvertasksEffPrice +
		\exprCertSplit*\NvertasksEffSplit + \exprCertLend*\NvertasksEffLend  	
		) / \NvertasksEff
	}%
	\FPround\tmp\tmp{2}\tmp
}

\newcommand{\covCertTot}{%
	\FPeval\tmp{
		(
		\covCertBank*\NvertasksEffBank + 
		\covCertVault*\NvertasksEffVault  + \covCertPrice*\NvertasksEffPrice +
		\covCertSplit*\NvertasksEffSplit + \covCertLend*\NvertasksEffLend  	
		) / \NvertasksEff
	}%
	\FPround\tmp\tmp{2}\tmp
}

\newcommand{\covGptFBank}{%
	\FPeval\tmp{
		(\NvertasksEffBank-\unBankFour) / (\NvertasksEffBank)*100
	}%
	\FPround\tmp\tmp{0}\tmp
}

\newcommand{\covGptFVault}{%
	\FPeval\tmp{
		(\NvertasksEffVault-\unVaultFour) / (\NvertasksEffVault)*100
	}%
	\FPround\tmp\tmp{2}\tmp
}

\newcommand{\covGptFPrice}{%
	\FPeval\tmp{
		(\NvertasksEffPrice-\unBetFour) / (\NvertasksEffPrice)*100
	}%
	\FPround\tmp\tmp{2}\tmp
}

\newcommand{\covGptFSplit}{%
	\FPeval\tmp{
		(\NvertasksEffSplit-\unSplitFour) / (\NvertasksEffSplit)*100
	}%
	\FPround\tmp\tmp{2}\tmp
}

\newcommand{\covGptFLend}{%
	\FPeval\tmp{
		(\NvertasksEffLend-\unLendFour) / (\NvertasksEffLend)*100
	}%
	\FPround\tmp\tmp{2}\tmp
}

\newcommand{\covGptFTot}{%
	\FPeval\tmp{
		(\NvertasksEff-(\unBankFour+\unVaultFour+\unBetFour+\unSplitFour+\unLendFour 
		)) / (\NvertasksEff)*100
	}%
	\FPround\tmp\tmp{2}\tmp
}


\newcommand{\DsCvlCertoraUNK}{5}
\newcommand{\DsCvlCertoraTP}{196}
\newcommand{\DsCvlCertoraTN}{178}
\newcommand{\DsCvlCertoraFP}{58}
\newcommand{\DsCvlCertoraFN}{57}
\newcommand{\DsCvlCertoraTot}{\the\numexpr \DsCvlCertoraTP + \DsCvlCertoraTN + \DsCvlCertoraFP + \DsCvlCertoraFN +\DsCvlCertoraUNK \relax}

\newcommand{\DsCvlCertoraAcc}{%
	\FPeval\tmp{ 
		(\DsCvlCertoraTP + \DsCvlCertoraTN)/(\DsCvlCertoraTP + \DsCvlCertoraTN+\DsCvlCertoraFP + \DsCvlCertoraFN)
		*100
	}%
	\FPround\tmp\tmp{0}\tmp
}

\newcommand{\DsCvlCertoraPrec}{%
	\FPeval\tmp{ 
		(\DsCvlCertoraTP)
		/
		(\DsCvlCertoraTP+\DsCvlCertoraFP)
		*100
	}%
	\FPround\tmp\tmp{0}\tmp
}

\newcommand{\DsCvlCertoraRec}{%
	\FPeval\tmp{ 
		(\DsCvlCertoraTP)
		/
		(\DsCvlCertoraTP+\DsCvlCertoraFN)
		*100
	}%
	\FPround\tmp\tmp{0}\tmp
}

\newcommand{\DsCvlCertoraSpec}{%
	\FPeval\tmp{ 
		(\DsCvlCertoraTN)
		/
		(\DsCvlCertoraTN+\DsCvlCertoraFP)
		*100	
	}%
	\FPround\tmp\tmp{0}\tmp
}

\newcommand{\DsCvlCertoraFOne}{%
	\FPeval\tmp{ 
		(2*\DsCvlCertoraTP)
		/
		(2*\DsCvlCertoraTP+\DsCvlCertoraFP + \DsCvlCertoraFN)
		*100
	}%
	\FPround\tmp\tmp{0}\tmp
}

\newcommand{\DsCvlGptCUNK}{0}
\newcommand{\DsCvlGptCTP}{240}
\newcommand{\DsCvlGptCTN}{215}
\newcommand{\DsCvlGptCFP}{22}
\newcommand{\DsCvlGptCFN}{17}

\newcommand{\DsCvlGptCTot}{\the\numexpr \DsCvlGptCTP + \DsCvlGptCTN + \DsCvlGptCFP + \DsCvlGptCFN + \DsCvlGptCUNK \relax}

\newcommand{\DsCvlGptCAcc}{%
	\FPeval\tmp{ 
		(\DsCvlGptCTP + \DsCvlGptCTN)/(\DsCvlGptCTP + \DsCvlGptCTN+\DsCvlGptCFP + \DsCvlGptCFN)
		*100
	}%
	\FPround\tmp\tmp{0}\tmp
}

\newcommand{\DsCvlGptCPrec}{%
	\FPeval\tmp{ 
		(\DsCvlGptCTP)
		/
		(\DsCvlGptCTP+\DsCvlGptCFP)
		*100
	}%
	\FPround\tmp\tmp{0}\tmp
}

\newcommand{\DsCvlGptCRec}{%
	\FPeval\tmp{ 
		(\DsCvlGptCTP)
		/
		(\DsCvlGptCTP+\DsCvlGptCFN)
		*100
	}%
	\FPround\tmp\tmp{0}\tmp
}

\newcommand{\DsCvlGptCSpec}{%
	\FPeval\tmp{ 
		(\DsCvlGptCTN)
		/
		(\DsCvlGptCTN+\DsCvlGptCFP)
		*100	
	}%
	\FPround\tmp\tmp{0}\tmp
}

\newcommand{\DsCvlGptCFOne}{%
	\FPeval\tmp{ 
		(2*\DsCvlGptCTP)
		/
		(2*\DsCvlGptCTP+\DsCvlGptCFP + \DsCvlGptCFN)
		*100
	}%
	\FPround\tmp\tmp{0}\tmp
}

\newcommand{\DsNoCvlGptCUNK}{0}
\newcommand{\DsNoCvlGptCTP}{63}
\newcommand{\DsNoCvlGptCTN}{97}
\newcommand{\DsNoCvlGptCFP}{9}
\newcommand{\DsNoCvlGptCFN}{4}

\newcommand{\DsNoCvlGptCTot}{\the\numexpr \DsNoCvlGptCTP + \DsNoCvlGptCTN + \DsNoCvlGptCFP + \DsNoCvlGptCFN + \DsNoCvlGptCUNK \relax}

\newcommand{\DsNoCvlGptCAcc}{%
	\FPeval\tmp{ 
		(\DsNoCvlGptCTP + \DsNoCvlGptCTN)/(\DsNoCvlGptCTP + \DsNoCvlGptCTN+\DsNoCvlGptCFP + \DsNoCvlGptCFN)
		*100
	}%
	\FPround\tmp\tmp{0}\tmp
}

\newcommand{\DsNoCvlGptCPrec}{%
	\FPeval\tmp{ 
		(\DsNoCvlGptCTP)
		/
		(\DsNoCvlGptCTP+\DsNoCvlGptCFP)
		*100
	}%
	\FPround\tmp\tmp{0}\tmp
}

\newcommand{\DsNoCvlGptCRec}{%
	\FPeval\tmp{ 
		(\DsNoCvlGptCTP)
		/
		(\DsNoCvlGptCTP+\DsNoCvlGptCFN)
		*100
	}%
	\FPround\tmp\tmp{0}\tmp
}

\newcommand{\DsNoCvlGptCSpec}{%
	\FPeval\tmp{ 
		(\DsNoCvlGptCTN)
		/
		(\DsNoCvlGptCTN+\DsNoCvlGptCFP)
		*100	
	}%
	\FPround\tmp\tmp{0}\tmp
}

\newcommand{\DsNoCvlGptCFOne}{%
	\FPeval\tmp{ 
		(2*\DsNoCvlGptCTP)
		/
		(2*\DsNoCvlGptCTP+\DsNoCvlGptCFP + \DsNoCvlGptCFN)
		*100
	}%
	\FPround\tmp\tmp{0}\tmp
}

\newcommand{\DsSolSolUNK}{48}
\newcommand{\DsSolSolTP}{72}
\newcommand{\DsSolSolTN}{95}
\newcommand{\DsSolSolFP}{14}
\newcommand{\DsSolSolFN}{25}

\newcommand{\DsSolSolTot}{\the\numexpr \DsSolSolTP + \DsSolSolTN + \DsSolSolFP + \DsSolSolFN + \DsSolSolUNK \relax}

\newcommand{\DsSolSolAcc}{%
	\FPeval\tmp{ 
		(\DsSolSolTP + \DsSolSolTN)/(\DsSolSolTP + \DsSolSolTN+\DsSolSolFP + \DsSolSolFN)
		*100
	}%
	\FPround\tmp\tmp{0}\tmp
}

\newcommand{\DsSolSolPrec}{%
	\FPeval\tmp{ 
		(\DsSolSolTP)
		/
		(\DsSolSolTP+\DsSolSolFP)
		*100
	}%
	\FPround\tmp\tmp{0}\tmp
}

\newcommand{\DsSolSolRec}{%
	\FPeval\tmp{ 
		(\DsSolSolTP)
		/
		(\DsSolSolTP+\DsSolSolFN)
		*100
	}%
	\FPround\tmp\tmp{0}\tmp
}

\newcommand{\DsSolSolSpec}{%
	\FPeval\tmp{ 
		(\DsSolSolTN)
		/
		(\DsSolSolTN+\DsSolSolFP)
		*100	
	}%
	\FPround\tmp\tmp{0}\tmp
}

\newcommand{\DsSolSolFOne}{%
	\FPeval\tmp{ 
		(2*\DsSolSolTP)
		/
		(2*\DsSolSolTP+\DsSolSolFP + \DsSolSolFN)
		*100
	}%
	\FPround\tmp\tmp{0}\tmp
}

\newcommand{\DsSolGptCUNK}{0}
\newcommand{\DsSolGptCTP}{110}
\newcommand{\DsSolGptCTN}{119}
\newcommand{\DsSolGptCFP}{19}
\newcommand{\DsSolGptCFN}{6}

\newcommand{\DsSolGptCTot}{\the\numexpr \DsSolGptCTP + \DsSolGptCTN + \DsSolGptCFP + \DsSolGptCFN \relax}

\newcommand{\DsSolGptCAcc}{%
	\FPeval\tmp{ 
		(\DsSolGptCTP + \DsSolGptCTN)/(\DsSolGptCTP + \DsSolGptCTN+\DsSolGptCFP + \DsSolGptCFN)
		*100
	}%
	\FPround\tmp\tmp{0}\tmp
}

\newcommand{\DsSolGptCPrec}{%
	\FPeval\tmp{ 
		(\DsSolGptCTP)
		/
		(\DsSolGptCTP+\DsSolGptCFP)
		*100
	}%
	\FPround\tmp\tmp{0}\tmp
}

\newcommand{\DsSolGptCRec}{%
	\FPeval\tmp{ 
		(\DsSolGptCTP)
		/
		(\DsSolGptCTP+\DsSolGptCFN)
		*100
	}%
	\FPround\tmp\tmp{0}\tmp
}

\newcommand{\DsSolGptCSpec}{%
	\FPeval\tmp{ 
		(\DsSolGptCTN)
		/
		(\DsSolGptCTN+\DsSolGptCFP)
		*100	
	}%
	\FPround\tmp\tmp{0}\tmp
}

\newcommand{\DsSolGptCFOne}{%
	\FPeval\tmp{ 
		(2*\DsSolGptCTP)
		/
		(2*\DsSolGptCTP+\DsSolGptCFP + \DsSolGptCFN)
		*100
	}%
	\FPround\tmp\tmp{0}\tmp
}

\newcommand{\DsNoSolGptCUNK}{0}
\newcommand{\DsNoSolGptCTP}{193}
\newcommand{\DsNoSolGptCTN}{193}
\newcommand{\DsNoSolGptCFP}{12}
\newcommand{\DsNoSolGptCFN}{15}

\newcommand{\DsNoSolGptCTot}{\the\numexpr \DsNoSolGptCTP + \DsNoSolGptCTN + \DsNoSolGptCFP + \DsNoSolGptCFN \relax}

\newcommand{\DsNoSolGptCAcc}{%
	\FPeval\tmp{ 
		(\DsNoSolGptCTP + \DsNoSolGptCTN)/(\DsNoSolGptCTP + \DsNoSolGptCTN+\DsNoSolGptCFP + \DsNoSolGptCFN)
		*100
	}%
	\FPround\tmp\tmp{0}\tmp
}

\newcommand{\DsNoSolGptCPrec}{%
	\FPeval\tmp{ 
		(\DsNoSolGptCTP)
		/
		(\DsNoSolGptCTP+\DsNoSolGptCFP)
		*100
	}%
	\FPround\tmp\tmp{0}\tmp
}

\newcommand{\DsNoSolGptCRec}{%
	\FPeval\tmp{ 
		(\DsNoSolGptCTP)
		/
		(\DsNoSolGptCTP+\DsNoSolGptCFN)
		*100
	}%
	\FPround\tmp\tmp{0}\tmp
}

\newcommand{\DsNoSolGptCSpec}{%
	\FPeval\tmp{ 
		(\DsNoSolGptCTN)
		/
		(\DsNoSolGptCTN+\DsNoSolGptCFP)
		*100	
	}%
	\FPround\tmp\tmp{0}\tmp
}

\newcommand{\DsNoSolGptCFOne}{%
	\FPeval\tmp{ 
		(2*\DsNoSolGptCTP)
		/
		(2*\DsNoSolGptCTP+\DsNoSolGptCFP + \DsNoSolGptCFN)
		*100
	}%
	\FPround\tmp\tmp{0}\tmp
}

\newcommand{\DsCvlGptQUNK}{6} 
\newcommand{\DsCvlGptQTP}{174}
\newcommand{\DsCvlGptQTN}{149}
\newcommand{\DsCvlGptQFP}{85}
\newcommand{\DsCvlGptQFN}{80}

\newcommand{\DsCvlGptQTot}{\the\numexpr \DsCvlGptQTP + \DsCvlGptQTN + \DsCvlGptQFP + \DsCvlGptQFN +\DsCvlGptQUNK \relax}

\newcommand{\DsCvlGptQAcc}{%
	\FPeval\tmp{ 
		(\DsCvlGptQTP + \DsCvlGptQTN)/(\DsCvlGptQTP + \DsCvlGptQTN+\DsCvlGptQFP + \DsCvlGptQFN)
		*100
	}%
	\FPround\tmp\tmp{0}\tmp
}

\newcommand{\DsCvlGptQPrec}{%
	\FPeval\tmp{ 
		(\DsCvlGptQTP)
		/
		(\DsCvlGptQTP+\DsCvlGptQFP)
		*100
	}%
	\FPround\tmp\tmp{0}\tmp
}

\newcommand{\DsCvlGptQRec}{%
	\FPeval\tmp{ 
		(\DsCvlGptQTP)
		/
		(\DsCvlGptQTP+\DsCvlGptQFN)
		*100
	}%
	\FPround\tmp\tmp{0}\tmp
}

\newcommand{\DsCvlGptQSpec}{%
	\FPeval\tmp{ 
		(\DsCvlGptQTN)
		/
		(\DsCvlGptQTN+\DsCvlGptQFP)
		*100	
	}%
	\FPround\tmp\tmp{0}\tmp
}

\newcommand{\DsCvlGptQFOne}{%
	\FPeval\tmp{ 
		(2*\DsCvlGptQTP)
		/
		(2*\DsCvlGptQTP+\DsCvlGptQFP + \DsCvlGptQFN)
		*100
	}%
	\FPround\tmp\tmp{0}\tmp
}

\newcommand{\DsNoCvlGptQUNK}{5}
\newcommand{\DsNoCvlGptQTP}{36}
\newcommand{\DsNoCvlGptQTN}{55}
\newcommand{\DsNoCvlGptQFP}{47}
\newcommand{\DsNoCvlGptQFN}{30}

\newcommand{\DsNoCvlGptQTot}{\the\numexpr \DsNoCvlGptQTP + \DsNoCvlGptQTN + \DsNoCvlGptQFP + \DsNoCvlGptQFN +\DsNoCvlGptQUNK \relax}

\newcommand{\DsNoCvlGptQAcc}{%
	\FPeval\tmp{ 
		(\DsNoCvlGptQTP + \DsNoCvlGptQTN)/(\DsNoCvlGptQTP + \DsNoCvlGptQTN+\DsNoCvlGptQFP + \DsNoCvlGptQFN)
		*100
	}%
	\FPround\tmp\tmp{0}\tmp
}

\newcommand{\DsNoCvlGptQPrec}{%
	\FPeval\tmp{ 
		(\DsNoCvlGptQTP)
		/
		(\DsNoCvlGptQTP+\DsNoCvlGptQFP)
		*100
	}%
	\FPround\tmp\tmp{0}\tmp
}

\newcommand{\DsNoCvlGptQRec}{%
	\FPeval\tmp{ 
		(\DsNoCvlGptQTP)
		/
		(\DsNoCvlGptQTP+\DsNoCvlGptQFN)
		*100
	}%
	\FPround\tmp\tmp{0}\tmp
}

\newcommand{\DsNoCvlGptQSpec}{%
	\FPeval\tmp{ 
		(\DsNoCvlGptQTN)
		/
		(\DsNoCvlGptQTN+\DsNoCvlGptQFP)
		*100	
	}%
	\FPround\tmp\tmp{0}\tmp
}

\newcommand{\DsNoCvlGptQFOne}{%
	\FPeval\tmp{ 
		(2*\DsNoCvlGptQTP)
		/
		(2*\DsNoCvlGptQTP+\DsNoCvlGptQFP + \DsNoCvlGptQFN)
		*100
	}%
	\FPround\tmp\tmp{0}\tmp
}

\newcommand{\DsSolGptQUNK}{1}
\newcommand{\DsSolGptQTP}{84}
\newcommand{\DsSolGptQTN}{84}
\newcommand{\DsSolGptQFP}{53}
\newcommand{\DsSolGptQFN}{32}

\newcommand{\DsSolGptQTot}{\the\numexpr \DsSolGptQTP + \DsSolGptQTN + \DsSolGptQFP + \DsSolGptQFN +\DsSolGptQUNK \relax}

\newcommand{\DsSolGptQAcc}{%
	\FPeval\tmp{ 
		(\DsSolGptQTP + \DsSolGptQTN)/(\DsSolGptQTP + \DsSolGptQTN+\DsSolGptQFP + \DsSolGptQFN)
		*100
	}%
	\FPround\tmp\tmp{0}\tmp
}

\newcommand{\DsSolGptQPrec}{%
	\FPeval\tmp{ 
		(\DsSolGptQTP)
		/
		(\DsSolGptQTP+\DsSolGptQFP)
		*100
	}%
	\FPround\tmp\tmp{0}\tmp
}

\newcommand{\DsSolGptQRec}{%
	\FPeval\tmp{ 
		(\DsSolGptQTP)
		/
		(\DsSolGptQTP+\DsSolGptQFN)
		*100
	}%
	\FPround\tmp\tmp{0}\tmp
}

\newcommand{\DsSolGptQSpec}{%
	\FPeval\tmp{ 
		(\DsSolGptQTN)
		/
		(\DsSolGptQTN+\DsSolGptQFP)
		*100	
	}%
	\FPround\tmp\tmp{0}\tmp
}

\newcommand{\DsSolGptQFOne}{%
	\FPeval\tmp{ 
		(2*\DsSolGptQTP)
		/
		(2*\DsSolGptQTP+\DsSolGptQFP + \DsSolGptQFN)
		*100
	}%
	\FPround\tmp\tmp{0}\tmp
}

\newcommand{\DsNoSolGptQUNK}{10}
\newcommand{\DsNoSolGptQTP}{126}
\newcommand{\DsNoSolGptQTN}{120}
\newcommand{\DsNoSolGptQFP}{79}
\newcommand{\DsNoSolGptQFN}{78}

\newcommand{\DsNoSolGptQTot}{\the\numexpr \DsNoSolGptQTP + \DsNoSolGptQTN + \DsNoSolGptQFP + \DsNoSolGptQFN  + \DsNoSolGptQUNK \relax}

\newcommand{\DsNoSolGptQAcc}{%
	\FPeval\tmp{ 
		(\DsNoSolGptQTP + \DsNoSolGptQTN)/(\DsNoSolGptQTP + \DsNoSolGptQTN+\DsNoSolGptQFP + \DsNoSolGptQFN)
		*100
	}%
	\FPround\tmp\tmp{0}\tmp
}

\newcommand{\DsNoSolGptQPrec}{%
	\FPeval\tmp{ 
		(\DsNoSolGptQTP)
		/
		(\DsNoSolGptQTP+\DsNoSolGptQFP)
		*100
	}%
	\FPround\tmp\tmp{0}\tmp
}

\newcommand{\DsNoSolGptQRec}{%
	\FPeval\tmp{ 
		(\DsNoSolGptQTP)
		/
		(\DsNoSolGptQTP+\DsNoSolGptQFN)
		*100
	}%
	\FPround\tmp\tmp{0}\tmp
}

\newcommand{\DsNoSolGptQSpec}{%
	\FPeval\tmp{ 
		(\DsNoSolGptQTN)
		/
		(\DsNoSolGptQTN+\DsNoSolGptQFP)
		*100	
	}%
	\FPround\tmp\tmp{0}\tmp
}

\newcommand{\DsNoSolGptQFOne}{%
	\FPeval\tmp{ 
		(2*\DsNoSolGptQTP)
		/
		(2*\DsNoSolGptQTP+\DsNoSolGptQFP + \DsNoSolGptQFN)
		*100
	}%
	\FPround\tmp\tmp{0}\tmp
}


\begin{abstract}
%
Ensuring the correctness of smart contracts is critical, as even subtle flaws can lead to severe financial losses. 
%
While bug detection tools able to spot common vulnerability patterns can serve as a first line of defense, most real-world exploits and losses stem from errors in the contract business logic.
Formal verification tools such as SolCMC and the Certora Prover address this challenge, but their impact remains limited by steep learning curves and restricted specification languages.
%
%
%
%
Recent works have begun to explore the use of large language models (LLMs) for  security-related tasks such as  vulnerability detection and test generation.
%
%
%
Yet, a fundamental question remains open: \change{can LLMs aid in assessing the validity of \emph{arbitrary} contract-specific properties?} 
%
%
%
%
In this paper, we provide the first systematic \change{empirical} evaluation of \gptC,
a state-of-the-art reasoning LLM, in this role. 
We benchmark its performance on a large dataset of verification tasks, compare its outputs against those of established formal verification tools,
and assess its practical effectiveness in real-world auditing scenarios.
Our study combines quantitative metrics with qualitative analysis, and shows that recent reasoning-oriented LLMs 
\change{--- although lacking soundness guarantees ---}
can be surprisingly effective {at predicting the (in)validity of complex properties}, suggesting a new frontier in the convergence of AI and formal methods for secure smart contract development and auditing.
\end{abstract}

\keywords{Smart Contracts \and Formal Verification \and LLMs \and Solidity}

\section{Introduction}
\label{sec:intro}

Since deployed smart contracts are immutable and often manage substantial economic assets, ensuring their correctness and security is paramount. Vulnerabilities and flaws in business logic have already resulted in losses exceeding \$6 billion~\cite{Chaliasos24icse}, underscoring the urgent need for safer and verifiable code.

For Solidity, the most adopted smart contract language, there exist lots of bug detection tools~\cite{Tolmach22csur,Zhang23icse}, 
which focus on finding common vulnerability patterns (\eg re-entrancy, overflows, locked ether).
However, 
it has been shown that the vast majority of the losses
due to real-world attacks are caused by logic errors in the contract code \cite{Chaliasos24icse}.
This motivates the need for formal verification tools, able to verify (or falsify) custom properties concerning the business logic of a contract. 
Several verification tools have been developed for this purpose, most notably SolCMC~\cite{solcmc-cav22-artifact}, shipped with the Solidity compiler, and the Certora Prover~\cite{certora}, a leading tool in the auditing industry.




While, in general, formal verification tools 
can significantly enhance the security of smart contracts, they also 
present several limitations: 
\begin{itemize}

\item properties need to be written in a \emph{formal} specification language, which 
\begin{enumerate}[label=(\alph*)]
\item requires a considerable manual effort,
\item restricts the set of expressible properties; 
\end{enumerate} 


\item for efficiency reasons, they often introduce unsound approximations of the smart contract semantics 
(\eg, they abstract from the gas mechanism, or neglect semantical corner cases such as coinbase and selfdestruct transactions);

\item despite such approximations, inherent issues of formal verification (\eg state explosion, undecidability of non-linear arithmetic) are unavoidable.
\end{itemize}

These limitations are not merely theoretical, but also have a practical impact on current Solidity verification tools.
Both SolCMC and the Certora Prover have been shown to be unsound (\ie,  they may report a property as verified when it is not)~\cite{BFMPS24fmbc}, and relevant classes of properties (\eg, strategic ones) remain beyond their scope~\cite{BCL25fmbc}.
Such shortcomings highlight the need for alternative techniques that can complement symbolic formal verification.


Recently, there has been a huge rise in works that adopt Large Language Models (LLMs)~\cite{brown2020language} 
for a variety of tasks in smart contract analysis,
including
common vulnerabilities detection~\cite{StillNeedManualAudit,SmartGuard,Liu2025BCRA,LLM-SmartAudit},
test generation~\cite{LLMtestGeneration}, and
specification generation~\cite{leite2024extracting,liu2025propertygpt}. 
%
While LLMs have been shown to perform reasonably well on such tasks,  
it remains unexplored whether they can also 
\change{play a role in the verification} of
custom, contract-specific properties. 
%
This challenge is considerably more complex, since the properties to verify can be arbitrary and are strictly tied to the specific contract under analysis.
Therefore, mere pattern recognition is not sufficient; a certain degree of reasoning capabilities is required. 
Recent models such as \mbox{\gptC} (released on August 7th, 2025) have claimed to show human-like reasoning capabilities; however, whether these capabilities  
transfer to smart contract verification has not been explored yet. 


If proven to be viable, the adoption of LLMs in the context of smart contracts verification  would be quite appealing.
%
Indeed, in contrast to formal verification tools, LLMs have the benefits of:
\begin{enumerate*}[label=(\roman*)]
	\item being able to work with arbitrary custom properties written in natural language,
    \item having a lower learning curve,
    \item 
    adopting a knowledge-based and pattern-driven problem-solving approach   that 
    complements
    the 
    symbolic
    reasoning approach used by formal methods~\cite{surveyReasoning}.
\end{enumerate*}
On the other hand, however, LLMs present significant downsides compared to formal verification: well-known weaknesses  include hallucinations, non-determinism, and lack of mathematical rigor.


%
%



 %


In this work, we investigate the viability of using LLMs for smart contract verification.
%
In particular, we consider the following research questions:
\begin{description}
	\item[RQ1)] Given a contract and a custom property written in natural language, are LLMs able to correctly determine whether the property holds or not?
	\item[RQ2)] Analysis of the output:
	\begin{itemize}
		\item[a)]
		   When LLMs correctly guess the truth value of a property, are they also able to produce a correct explanation or  counterexample that substantiates their answer?
		\item[b)] When LLMs  incorrectly guess   the truth value of a property, what are the causes of failure?
	\end{itemize}
	\item[RQ3)] What are the strengths and weaknesses of LLMs compared to mainstream Solidity verification tools?
\end{description}

To rigorously address these research questions, it is essential to operate in a controlled environment where the ground truth of the considered properties can be manually established.
%
%
To further investigate whether LLMs can be useful 
\emph{in the wild}, 
we consider a fourth research question:
%
%
%
\begin{description}
    \item[RQ4)] Can LLMs be effectively used in real-world
    auditing scenarios?
\end{description}


This paper addresses the previous research questions by providing the following key contributions:
\begin{itemize}

\item For \textbf{RQ1-3}, we constructed the largest public dataset for Solidity verification, to the best of our knowledge. 
The dataset defines \Nvertasks\xspace verification tasks distributed across $\Nusecases$ paradigmatic use cases, each one implemented in several mutations containing subtle bugs or alternative behaviours.
The dataset contains a wide range of properties, encompassing function specs,  state and transition invariants, and also complex strategic properties that are beyond the scope of existing symbolic verification tools.

\item For \textbf{RQ1}, we evaluated \gptQ and \gptC over our dataset. 
Results show that \gptC consistently achieves over $85\%$ across all prediction metrics and use cases, outperforming \gptQ by a margin of $25\%-30\%$. 

\item For \textbf{RQ2}, we manually analysed the explanations and counterexamples produced by the LLMs to justify their classifications.
This analysis confirmed that both models properly interpreted the natural-language properties. 
We then assessed the correctness (\ie, absence of logical errors) and completeness of the explanations, finding a clear correlation between \gptC's high prediction metrics and its stronger reasoning capabilities. In contrast, \gptQ often produced vague or erroneous arguments leading to hallucinations.

\item For \textbf{RQ3}, we compared LLMs and the state-of-the-art symbolic verification tools \solcmc and \certora, finding that \gptC achieves higher prediction metrics than the two tools on their respective 
expressibility sub-datasets,
also behaving consistently well on the set of the (often more complex) properties not expressible in the two tools. 

\item For \textbf{RQ4}, we considered Safe and InfiniFi, two projects recently audited by Certora~\cite{CertoraReports}.
We ran \gptC over the properties addressed in the two reports, showcasing the ability of \gptC to correctly classify most of them.
We further demonstrated how \gptC enhances verification auditing
by detecting inconsistencies between property descriptions and formal specifications,  
and substantiating violations by producing detailed counterexample traces.


\end{itemize}

Overall, our analysis shows that recent models such as \gptC can be effectively used to support smart contract auditing.
To foster reproducibility, we release our dataset, the scripts needed to run the experiments, and the complete experimental results in a public repository~\cite{anonymous_repo}.
\section{Background}
\label{sec:background}

This work lies at the intersection of formal verification and AI–based code analysis. We provide below the an essential background on these topics.

\subsection{Symbolic verification tools for Solidity}
\label{sub:background_tools}

Several verification tools for Solidity have been proposed in recent years (see~\Cref{sec:related}).
Among them, two have achieved industrial adoption:
SolCMC~\cite{Solcmc}, a symbolic model checker integrated with the Solidity compiler, and the Certora Prover~\cite{certora,certora-wip}, one of the leading tools in security auditing of smart contracts.

In SolCMC, properties are expressed through assertions embedded in Solidity code. 
These assertions specify state invariants that must hold in all reachable contract states.
SolCMC translates the instrumented contract into a set of logical constraints, 
which are then fed to a SMT/CHC solver (Z3~\cite{DeB08} or Eldarica~\cite{Eldarica}).

Unlike SolCMC, Certora decouples the specification of the properties from the contract code, providing a domain-specific language (CVL) for specs.
A CVL spec defines constraints on the execution of a contract, and assertions that must be true in all states satisfying the given constraints.   
Certora compiles the Solidity contract and the CVL spec into a logical formula, and relies on SMT solvers to determine if the spec is satisfied in all contract states.

Both tools make some approximations of the Solidity semantics (\eg, both of them abstract from gas costs) to make verification more practical. 
These approximations, however, may yield false negatives (\eg, a property holds but the tool classifies it as violated), or false positives (the property is violated, but the tool reports it as satisfied).
Assessing the reliability of a classification then requires a deep expertise on both the Solidity semantics and the specific approximations made by the tool, which in practice leads to a costly, iterative process of refining the specifications.  


\subsection{Large Language Models for reasoning}
\label{subesec:reasoning}
Large Language Models are trained to predict the next token in a sequence, effectively capturing statistical regularities from massive text corpora. At their core, these models do not reason in the human sense: they generate continuations based on learned patterns rather than explicit symbolic inference. Nevertheless, suitable prompting strategies can produce behaviors that resemble logical or step-wise reasoning, enabling models to address multi-step problems beyond surface-level pattern matching. In this context, reasoning refers to the model’s ability to decompose a task into intermediate steps, evaluate partial conclusions, and synthesize them into a final answer.
A consistent finding is that prompting models to think in steps improves accuracy on multi-step math, logic, and symbolic tasks. In particular, \textit{Chain-of-Thought} (CoT) prompting, eliciting brief intermediate rationales, boosts performance across arithmetic, commonsense, and symbolic reasoning benchmarks, especially for larger models~\cite{wei2022chain}.

A closely related variant 
is \textit{Zero-shot CoT}, which adds a simple instruction to reason step-by-step without providing exemplars. This single instruction has been shown to markedly improve zero-shot results on diverse reasoning benchmarks~\cite{kojima2022large}. Prior work shows that promising results can be achieved by prompting the model with a Zero-shot CoT with hidden rationales (i.e., the reasoning is performed but not revealed verbatim): this step-wise prompting can improve reliability without requiring few-shot demonstrations~\cite{kojima2022large}.

\section{Dataset design}
\label{sec:dataset}

Our research questions require both a systematic analysis (RQ1–3) and coverage of real-world scenarios (RQ4). Following common practice~\cite{StillNeedManualAudit,Guo2024,LLM-SmartAudit}, we therefore combine complementary methodologies. 
For RQ1–3, we construct a controlled dataset of paradigmatic use cases and verification tasks, for which the ground truth can be manually established.
We describe in this section the design of the dataset, 
in the following section the methodology used to perform the experiments for RQ1-3, and in \Cref{sec:RQ4} the related discussion for RQ4.

\begin{table}[b]
    \centering
    \caption{Dataset metrics (lines of code averaged over number of versions).}
    \label{tab:dataset}
    \begin{tabular}{|c|c|c|c|l|}
    \hline
    \textbf{Use case} & \textbf{LoC} & \textbf{\#Mut} & \textbf{\#Prop} & \hspace{40pt}\textbf{Solidity features} \\
    \hline
    Bank & \NlBank & \NvBank & \NpBank & ETH transfer, mappings, linear arith. 
    \\
    Vault & \NlVault & \NvVault & \NpVault & ETH transfer, time, state machine 
    \\
    PriceBet & \NlPriceBet & \NvPriceBet & \NpPriceBet & ETH transfer, external oracle 
    \\
    PaymentSplitter & \NlPaymentSplitter & \NvPaymentSplitter & \NpPaymentSplitter & ETH transfer, loops, non-linear arith. 
    \\
    LendingProtocol & \NlLP & \NvLP & \NpLP & 
    ERC20 tokens, mappings, non-linear arith.
    \\
    \hline
    \end{tabular}
\end{table}


Our dataset comprises \Nusecases\xspace use cases, covering different Solidity features and different levels of complexity (\Cref{tab:dataset}):
\begin{enumerate}
\item \emph{Bank}: this is a minimal wallet contract that allows users to deposit and withdraw ETH. Its simplicity allows us to design properties whose truth may depend on subtle aspects of the Solidity semantics, \eg contract-to-contract calls, gas, reentrancy, and \emph{selfdestruct} transactions.    
The properties in the dataset include \emph{function specs} (\eg, a non-reverting call to a certain function produces a certain effect on caller and callee), \emph{state invariants} (\eg, the sum of all credits does not exceed the contract balance), \emph{metamorphic properties} (\eg, two different sequences of transactions produce the same effect on the contract state), and \emph{strategic properties} (\eg, there exists a sequence of transactions that produce some desired effect on the contract state --- such as no ETH remains frozen in the contract).
 
\item \emph{Vault}: this is a contract allowing the owner to deposit and withdraw ETH, with a mechanism to cancel pending withdrawal requests through a recovery key. Its implementations require time constraints and an automaton-like behaviour (\ie, certain actions can only be performed in certain states).
The properties now also predicate about the immutability of contract variables (\eg, a certain address is the same in every contract state --- or even within the execution of a function), and liveness properties with fairness assumptions (\eg, ``if $A$ knows both the owner and recovery key, and no one else knows the recovery key, then in every fair trace $A$ is able to eventually transfer any fraction of the contract balance, while no $B \neq A$ can do the same''). 

\item \emph{PriceBet}: this is a two-players bet on the price of a token, which is queried from an external oracle~\cite{Babel23clockwork}.
The properties in the dataset explore additional aspects such as assumptions about external contracts and MEV attacks (\eg, an adversary can frontrun another player's transaction to win the bet). 

\item \emph{PaymentSplitter}: this is an OpenZeppelin template that allows a set of shareholders to pull payments from a contract proportionally to their shares~\cite{paymentsplitter}. It allows us to explore the impact of non-linear arithmetic (used to determine the releasable ETH for each payee) and loops (used in the constructor to initialize the shareholders data).

\item \emph{LendingProtocol}: this is the most complex contract in our controlled dataset, as it implements a subset of the functionality of the Aave protocol~\cite{aaveimpl} (with some simplifications in order to keep the ground truth reliable).
It allows us to explore non-linear arithmetic, integer truncations, and economic properties (\eg, certain actions produce certain gains for the involved users\cite{BL25arxiv}). 

\end{enumerate}

The mutations of each use case are constructed manually, by introducing logic errors or divergent behaviours that may affect the validity of some the given properties.  
We provide each use case with a set of properties, written in natural language, which are common to all the mutations.
Some properties are given in slightly different versions, to assess how verification is affected by different assumptions on the context (\eg, sometimes we assume that an address has a \lstinline{receive} function that just accepts all ETH, in order to verify if ETH transfers work as intended at least in this simple case). 

For each verification task --- \ie, pair (property, mutation) ---
we provide the \emph{ground truth}, \ie a boolean telling whether the property holds or not for that mutation.
The first three use cases (\emph{Bank}, \emph{Vault} and \emph{PriceBet}) are simple enough for us to enable us to construct this ground truth manually (despite the apparent simplicity, however, there are several cases for which a certain degree of expertise is required, \eg when the ground truth is affected by quirky semantical aspects such as reentrancy, \emph{selfdestruct} transactions, or gas passed along with low-level calls).
For the most complex use cases (\emph{PaymentSplitter} and \emph{LendingProtocol}), besides manual analysis, we have consolidated the ground truth by providing Hardhat PoCs for the verification tasks where the property is violated 
(the code of these PoCs is in the dataset).
The verification tasks for which we were not sure about the ground truth are not included in the experiments.

There are two key remarks in the design of properties. 
First, we carefully crafted the natural language specifications to be as precise as possible, thereby reducing the risk of LLMs giving wrong answers  because of misunderstandings. 
The qualitative analysis of the LLMs outputs (in particular, the coherency metrics, see~\Cref{sec:RQ2}) shows that the LLMs understood our specifications with high confidence. 
Second, we ensured a balanced range of complexity across properties, avoiding biases introduced by verification tasks that are either too easy or too difficult. 
Moreover, our dataset includes both properties expressible by symbolic verification tools (\eg, function specs, state and transition invariants) and properties beyond their expressive power, allowing us to assess the effectiveness of LLMs on tasks that remain out of reach for existing symbolic verification tools.

\paragraph{Symbolic verification tools.}

To compare LLMs against symbolic verification tools, we focussed on \solcmc and \certora, for the reasons outlined in~\Cref{sec:background}.
%
For this purpose, we translated the natural-language properties into formal specifications for both tools and included them in the dataset.
Such encodings were not always feasible: \solcmc is limited to state invariants, while CVL cannot express certain classes of properties, such as strategic ones.   
Hence, comparison with \gptC were performed on the relevant subsets of the dataset (see~\Cref{sec:RQ3}). 

A crucial point here is that writing these formal specifications was not a one-shot process, but an iterative one.
In particular, even when a property was true and its initial specification was a faithful encoding of the natural-language statement, the tool often failed to verify it.
This issue arose more frequently with \certora, for two main reasons.
First, \certora uses conservative over-approximations of contract behaviour, sometimes considering cases that are impossible in practice.
For example, it may assume that a contract can call itself as a sender, even when such behavior is excluded by the code. 
In these situations, we had to introduce additional technical assumptions to exclude spurious behaviors that falsify the property. 
Second, some true properties require auxiliary invariants that the tool cannot automatically infer. In these cases, we supplied the missing invariants as part of the specification, effectively serving as proof hints to guide the verifier toward establishing the property.


%


\section{Experimental setup for LLMs evaluation}
\label{sec:methodology}

The primary focus of this study is \gptC, the most recent LLM released by OpenAI, which incorporates a dedicated reasoning module (GPT-5 Thinking) designed to address complex problems. 
As a baseline for comparison, we use \gptQ, which lacks such a reasoning component. 
To evaluate both models, we developed a Python script that takes as input a set of verification tasks, queries the LLMs via the OpenAI APIs, and collects their outputs for analysis. 

Each query is constructed using a carefully crafted prompt that follows a zero-shot, chain-of-thought style. 
The prompt (see~\Cref{app:prompt}) requires the model to: 
\begin{inlinelist}
\item determine whether a given property holds across all reachable contract states; 
\item explain the reasoning process leading to this decision; and 
\item if the property does not hold, provide a counterexample in the form of a transaction sequence that violates it. 
\end{inlinelist}
We additionally allow the LLMs to respond with ``unknown'' when uncertain, although in practice we observed that both \gptQ and \mbox{\gptC} rarely exercised this option.
To ensure a fair comparison between LLMs and formal verification tools in RQ3, we made explicit in the prompt certain common abstractions: in particular, we instructed GPT to neglect gas costs: \eg, when an address $A$ sends a transaction to withdraw 1 ETH from the contract, we want GPT to assume that $A$'s balance increases by exactly 1 ETH, without accounting for the gas spent be $A$ to execute the transaction. 
This abstraction is consistent with the behaviour observed both in \solcmc and \certora. 

To mitigate bias toward simpler tasks, we did not feed the models with the entire dataset. 
Rather, we sampled verification tasks from the three simpler use cases, while retaining the complete set of tasks from the others.
The resulting sub-dataset (hereafter, referred to as \dataset) contains \NvertasksEff\xspace verification tasks, and it is nearly balanced, featuring $\sim\mbox{1:1}$ ratio between simpler and complex use cases, and a comparable
$\sim\mbox{1:1}$ ratio between positive and negative tasks.
\change{
Text generation used the API's default decoding parameters at the time of the experiments. Specifically, the sampling temperature was set to 1.0, nucleus sampling was disabled (top-p = 1.0), and no frequency or presence penalties were applied. The maximum number of generated tokens was 30,000, with generation terminating either upon reaching this limit or according to the model's internal end-of-sequence criteria. The total cost of the experiments was below €100. The full dataset—including scripts, formal specifications in SolCMC and Certora, and all LLM outputs, is available in~\cite{anonymous_repo}.}
%








\section{RQ1: LLMs metrics on the controlled dataset}
\label{sec:RQ1}
\begin{table}[b]
	\centering
	\caption{Quantitative analysis for \gptQ (in gray) and \gptC (in cyan).}
    \label{table:resRQ1}
    \resizebox{\textwidth}{!}{
	\begin{tabular}{|l|r|r|r|r|r|r|r|r|r|r|r|}
		\hline
		Use case & Tot. & UNK & TP & TN & FP & FN & Accuracy & Precision & Recall & Specificity & F1 Score \\ \hline
		\hline

		\rowgptQ Bank & \anBankFour & \unBankFour & \TPBankFour & \TNBankFour & \FPBankFour & \FNBankFour & \acBankFour & \prBankFour & \reBankFour & \spBankFour & \fsBankFour \\ \hline
        
		\rowgptQ Vault & \anVaultFour & \unVaultFour & \TPVaultFour & \TNVaultFour & \FPVaultFour & \FNVaultFour & \acVaultFour & \prVaultFour & \reVaultFour & \spVaultFour & \fsVaultFour \\ \hline
		
        \rowgptQ PriceBet & \anBetFour & \unBetFour & \TPBetFour & \TNBetFour & \FPBetFour & \FNBetFour & \acBetFour & \prBetFour & \reBetFour & \spBetFour & \fsBetFour \\ \hline
		
		\rowgptQ PaymentSplitter & \anSplitFour & \unSplitFour & \TPSplitFour & \TNSplitFour & \FPSplitFour & \FNSplitFour & \acSplitFour & \prSplitFour & \reSplitFour & \spSplitFour & \fsSplitFour \\ \hline
		
		\rowgptQ LendingProtocol & \anLendFour & \unLendFour & \TPLendFour & \TNLendFour & \FPLendFour & \FNLendFour & \acLendFour & \prLendFour & \reLendFour & \spLendFour & \fsLendFour \\ \hline

        \rowgptQ OVERALL (\gptQ) & \anOverallFour & \unOverallFour & \TPOverallFour & \TNOverallFour & \FPOverallFour & \FNOverallFour & \acOverallFour & \prOverallFour & \reOverallFour & \spOverallFour & \fsOverallFour \\ \hline
        \noalign{\vskip 1mm}  \hline
        
		\rowgptC Bank & \anBankFive & \unBankFive & \TPBankFive & \TNBankFive & \FPBankFive & \FNBankFive & \acBankFive & \prBankFive & \reBankFive & \spBankFive & \fsBankFive \\ \hline
        
		\rowgptC Vault & \anVaultFive & \unVaultFive & \TPVaultFive & \TNVaultFive & \FPVaultFive & \FNVaultFive & \acVaultFive & \prVaultFive & \reVaultFive & \spVaultFive & \fsVaultFive \\ \hline
		
        \rowgptC PriceBet & \anBetFive & \unBetFive & \TPBetFive & \TNBetFive & \FPBetFive & \FNBetFive & \acBetFive & \prBetFive & \reBetFive & \spBetFive & \fsBetFive \\ \hline
		
		\rowgptC PaymentSplitter & \anSplitFive & \unSplitFive & \TPSplitFive & \TNSplitFive & \FPSplitFive & \FNSplitFive & \acSplitFive & \prSplitFive & \reSplitFive & \spSplitFive & \fsSplitFive \\ \hline
		
		\rowgptC 
		LendingProtocol & \anLendFive & \unLendFive & \TPLendFive & \TNLendFive & \FPLendFive & \FNLendFive & \acLendFive & \prLendFive & \reLendFive & \spLendFive & \fsLendFive \\ \hline
        
		\rowgptC OVERALL (\gptC) & \anOverallFive & \unOverallFive & \TPOverallFive & \TNOverallFive & \FPOverallFive & \FNOverallFive & \acOverallFive & \prOverallFive & \reOverallFive & \spOverallFive & \fsOverallFive \\ \hline 
	\end{tabular}
    } 
\end{table}

\Cref{table:resRQ1} reports the experimental results of \gptC and \gptQ on dataset \dataset.
\gptC achieves an F1 score above $\fsSplitFive$ in all use cases ($92\%$ overall), outperforming \gptQ on all of them with a performance gap of $25\%-30\%$ on every prediction metric. 
This result aligns with prior findings on \gptQ's limited logical reasoning abilities~\cite{EvaluatingLogicalReasoningAbilityGPT}.
The most challenging use case was \emph{PaymentSplitter}, where \gptQ performs only slightly better than random guessing, while \mbox{\gptC} still maintains performances above 80\% across all metrics.
We observe that \mbox{\gptC} has a very low rate of FP, which is a  desirable soundness property of verification tools.
We elaborate in~\Cref{sec:RQ2} the main causes of FPs and FNs.

We were quite surprised by \gptC's performance on \emph{LendingProtocol}, the most complex use case in our dataset.
We elaborated some explanations for this result (see also~\Cref{sec:RQ2}).
First, to address its complexity, we were more conservative than in the other use cases when crafting the mutations, to minimize the risk of errors in the ground truth.
So, here the mutations do not exhibit the diversity of behaviours that we reached in the other use cases: this results in several properties sharing the same truth for most mutations. 
For the same reason, several properties we devised for \emph{LendingProtocol}  are function specs for the various operations (which are easire to check for the ground truth), while there are less strategic properties, which are more difficult to assess manually, and arguably also for the LLMs.
Additionally, this use case deals with tokens instead of ETH: to avoid dealing with malicious token implementations, in many properties we added the assumption that they are standard ERC20 tokens: this ruled out reentrancy, which was one of the causes of failure of \gptC in  other use cases.  
We also observed that \gptC is very good at reasoning about non-linear integer arithmetic, which was required to solve many of the verification tasks. 






\section{RQ2: Analysis of LLMs explanations}
\label{sec:RQ2}

\begin{table}[b]
	\centering
    \small
    \caption{Qualitative analysis. In gray, results for \gptQ on the Bank use case. In cyan, results for \gptC on all use cases.}
    \label{table:resRQ2}
	\begin{tabular}{|l|r|r|r|r|r|r|r|r|}
		\hline
		 & \multicolumn{4}{c|}{TP + TN} & \multicolumn{4}{c|}{FP + FN} \\ \hline
		Use case & \#  & Correct & Complete & Coherent & \#  & Correct & Complete & Coherent \\ 
		\hline \noalign{\vskip 1mm}  \hline 		
		\rowgptQ
		 Bank
		& \anaTrueBankFour & \corTrueBankFourgr & \comTrueBankFourgr & \cohTrueBankFourgr & \anaFalseBankFour & \corFalseBankFourgr & \comFalseBankFourgr & \cohFalseBankFourgr \\ \hline
		 \noalign{\vskip 1mm}  \hline 
		\rowgptC Bank   & \anaTrueBankFive & \corTrueBankFivegr & \comTrueBankFivegr & \cohTrueBankFivegr & \anaFalseBankFive & \corFalseBankFivegr & \comFalseBankFivegr & \cohFalseBankFivegr \\ \hline
        
		\rowgptC Vault & \anaTrueVaultFive & \corTrueVaultFivegr & \comTrueVaultFivegr & \cohTrueVaultFivegr & \anaFalseVaultFive & \corFalseVaultFivegr & \comFalseVaultFivegr & \cohFalseVaultFivegr \\ \hline
		
        \rowgptC PriceBet & \anaTrueBetFive & \corTrueBetFivegr & \comTrueBetFivegr & \cohTrueBetFivegr & \anaFalseBetFive & \corFalseBetFivegr & \comFalseBetFivegr & \cohFalseBetFivegr \\ \hline
		
        \rowgptC PaymentSplitter & \anaTrueSplitFive & \corTrueSplitFivegr & \comTrueSplitFivegr & \cohTrueSplitFivegr & \anaFalseSplitFive & \corFalseSplitFivegr & \comFalseSplitFivegr & \cohFalseSplitFivegr \\ \hline
		
        \rowgptC LendingProtocol & \anaTrueLendFive & \corTrueLendFivegr & \comTrueLendFivegr & \cohTrueLendFivegr & \anaFalseLendFive & \corFalseLendFivegr & \comFalseLendFivegr & \cohFalseLendFivegr \\ \hline
		
		
	\end{tabular}
\end{table}

We now study the quality of the explanations and of the counterexamples provided by the LLMs.
We manually analyzed answers of \gptQ and \gptC (to, respectively, \anaTrueOverallFour\xspace and \anaTrueOverallFive\xspace tasks),
evaluating them under three criteria: \emph{coherence}, \emph{completeness}, and {\emph{correctness}}.
By {coherence}, we mean that the property has been properly understood by the LLM: we assess this by checking that the conclusions indeed imply that the property is verified/falsified, and that, if some hypotheses have been used, they match the ones in the property.
By correctness, we mean that the explanation does not contain logical errors: 
for \True-answers, each implication ``premise $\implies$ conclusion'' must be indeed logically valid; for \False-answers, we additionally check that each step of  the counterexample is executable in the given states.
By completeness, we mean, 
for \True-answers, that the explanation uses all the assumption mentioned in the property that are necessary for it to hold; 
for \False-answers, that the counterexample contains all the necessary steps. 
Overall, correct and complete counterexamples can be translated into  executable PoCs.

	
	
	
For each answer, we mark each criterion with $0$ or $1$,
and add annotations whenever necessary.
\Cref{table:resRQ2} reports the results. 
%
Although this analysis necessarily present a certain degree of subjectivity,  it helps better understanding how faithfully the metrics for RQ1 represent the reasoning capabilities of LLMs.


 %
	%
    %
	
To compare \gptQ \emph{vs.} \gptC, we focussed on the Bank contract, on which we exhaustively evaluated all the answers. 
Coherence is rarely an issue, 
suggesting that both models are able to correctly interpret the property; this is further supported by the fact that slight variations of the properties (\eg, different assumptions) often lead to the expected explanations and counterexamples.
The main cause of false results for \gptQ is incorrectness.
\gptC, on the contrary, performs reasonably well on all criteria. 
For wrong answers, the cause distributes rather equally among the three criteria;
for correct answers, the main weak point is completeness.
\change{Since the simplest use case already shows a significant gap in reasoning capabilities between the two models, we did not extend the qualitative comparison of GPT-4's and GPT5's answers to the other use cases.}

To address RQ2b, we analyzed the causes of incorrect answers. 
By manually inspecting the verification tasks that produced FPs, we identified a few recurring issues: 
\begin{inlinelist}
\item low-level calls triggering ETH transfers; 
\item \emph{selfdestruct} transactions that increase a contract's balance;
\item integer overflows (primarily, in \emph{PaymentSplitter}). 
\end{inlinelist}
We note, however, that \gptC is not consistently incorrect on these features, as it produces correct answers in most cases. 
A common cause of FNs is the use of \lstinline{transfer} to send ETH: here, \gptC often misses that this operation does not carry enough gas to allow the callee to perform a \emph{selfdestruct}. 




\section{RQ3: LLMs \emph{vs.} formal verification tools}
\label{sec:RQ3}
\begin{table}[!b]
    \centering
    \small
    \caption{Expressibility (left) and Coverage (right, in parentheses) percentages.
    }
    \label{table:Coverage}
	\begin{center}
		\begin{tabular}{|l|l|\columnSol|\columnCert|}
			\hhline{|-|-|-|-|}
			Use case& \# & SolCMC & Certora  \\ 
			\hhline{|-|-|-|-|}
			\hhline{|=|=|=|=|}
			{Bank} & {\NvertasksEffBank} & \exprcov{\exprSolBank}{\covSolBank} & \exprcov{\exprCertBank}{\covCertBank}  \\ 
			\hhline{|-|-|-|-|}
			{Vault} & {\NvertasksEffVault} &  \exprcov{\exprSolVault}{\covSolVault} & \exprcov{\exprCertVault}{\covCertVault}  \\ 
			\hhline{|-|-|-|-|}
			{PriceBet} & {\NvertasksEffPrice} &  \exprcov{\exprSolPrice}{\covSolPrice} & \exprcov{\exprCertPrice}{\covCertPrice} \\ 
			\hhline{|-|-|-|-|}			
			{PaymentSplitter} & {\NvertasksEffSplit} &  \exprcov{\exprSolSplit}{\covSolSplit} & \exprcov{\exprCertSplit}{\covCertSplit} \\ 
			\hhline{|-|-|-|-|}
			LendingProtocol
			& {\NvertasksEffLend} &  \exprcov{\exprSolLend}{\covSolLend} & \exprcov{\exprCertLend}{\covCertLend}  \\ 
			\hhline{|-|-|-|-|}
			\hhline{|=|=|=|=|}
					{OVERALL} & {\NvertasksEff} &  \exprcov{\exprSolTot}{\covSolTot} & \exprcov{\exprCertTot}{\covCertTot}  \\ 
					\hhline{|-|-|-|-|}
		\end{tabular}
	\end{center}
\end{table}

%
%
%
%
As discussed in \Cref{sec:dataset}, only certain properties can be expressed in 
the specification languages of \solcmc and \certora.
We define \emph{expressibility} as the percentage of verification tasks that can be expressed in the given tool.
We also measure the \emph{coverage}, \ie the percentage of tasks for which the tool returns \True or \False 
(instead of \Unknown or timeouts).
%

\Cref{table:Coverage} displays these metrics.
%
We see that \certora is significantly more expressive than \solcmc,
being able to express almost $75\%$ of the tasks (\emph{vs}.~$35\%$).
\solcmc struggles on complex use cases: 
for \emph{PaymentSplitter}, its covers $<50\%$ of the expressible tasks, while its expressiveness on \emph{LendingProtocol} is negligible.
\certora covers all expressible tasks except for \emph{LendingProtocol}, where it times out on ${\sim}5\%$ experiments.
%


%
To ensure a fair comparison of the LLMs metrics in \Cref{sec:RQ1}, which cover the entire dataset \dataset, with those of SolCMC and Certora, 
we define two sub-datasets corresponding to the expressible verification tasks of the two tools: we refer to them as \datasetSolCMC and \datasetCVL, respectively.
%
%
To study how LLMs perform on the  tasks not expressible by \solcmc and \certora, we define the complementary sub-datasets
${\datasetNoSolCMC := \dataset \setminus \datasetSolCMC}$  
and
${\datasetNoCVL := \dataset \setminus \datasetCVL}$.

\begin{table}[t]
	\centering
    \small
    \caption{Results by sub-datasets (all use cases).
    	\label{table:resBySubDSCVL}}
    \resizebox{\textwidth}{!}{
	\begin{tabular}{|ll|c|c|c|c|c|c|c|c|c|c|c|}
		\hline
		Dataset & Tool & Tot. & UNK & TP & TN & FP & FN & Accuracy & Precision  & Recall  & Specificity & F1 score \\ \hline
		\noalign{\vskip 1mm} 
		\hline
		
	 \rowCert	\datasetCVL &  Certora & \DsCvlCertoraTot & \DsCvlCertoraUNK & \DsCvlCertoraTP & \DsCvlCertoraTN &\DsCvlCertoraFP & \DsCvlCertoraFN & \DsCvlCertoraAcc\% & \DsCvlCertoraPrec\% & \DsCvlCertoraRec\% & \DsCvlCertoraSpec\% & \DsCvlCertoraFOne\% \\ \hline
	 
	\rowgptC	\datasetCVL & \gptC & \DsCvlGptCTot & \DsCvlGptCUNK & \DsCvlGptCTP & \DsCvlGptCTN &\DsCvlGptCFP & \DsCvlGptCFN &  \DsCvlGptCAcc\% & \DsCvlGptCPrec\% & \DsCvlGptCRec\% & \DsCvlGptCSpec\% & \DsCvlGptCFOne\%  \\ \hline 
	
	\rowgptC	\datasetNoCVL & \gptC & \DsNoCvlGptCTot & \DsNoCvlGptCUNK & \DsNoCvlGptCTP & \DsNoCvlGptCTN &\DsNoCvlGptCFP & \DsNoCvlGptCFN &  \DsNoCvlGptCAcc\% & \DsNoCvlGptCPrec\% & \DsNoCvlGptCRec\% & \DsNoCvlGptCSpec\% & \DsNoCvlGptCFOne\%  \\ \hline 
	
	\rowgptQ	\datasetCVL & \gptQ & \DsCvlGptQTot & \DsCvlGptQUNK & \DsCvlGptQTP & \DsCvlGptQTN &\DsCvlGptQFP & \DsCvlGptQFN &  \DsCvlGptQAcc\% & \DsCvlGptQPrec\% & \DsCvlGptQRec\% & \DsCvlGptQSpec\% & \DsCvlGptQFOne\%  \\ \hline 
	
	\rowgptQ	\datasetNoCVL & \gptQ & \DsNoCvlGptQTot & \DsNoCvlGptQUNK & \DsNoCvlGptQTP & \DsNoCvlGptQTN &\DsNoCvlGptQFP & \DsNoCvlGptQFN &  \DsNoCvlGptQAcc\% & \DsNoCvlGptQPrec\% & \DsNoCvlGptQRec\% & \DsNoCvlGptQSpec\% & \DsNoCvlGptQFOne\%  \\ \hline 
	
		\noalign{\vskip 1mm} 
		\hline
		\rowSol	\datasetSolCMC &  SolCMC & \DsSolSolTot & \DsSolSolUNK & \DsSolSolTP & \DsSolSolTN &\DsSolSolFP & \DsSolSolFN & \DsSolSolAcc\% & \DsSolSolPrec\% & \DsSolSolRec\% & \DsSolSolSpec\% & \DsSolSolFOne\% \\ \hline
		
		\rowgptC	\datasetSolCMC & \gptC & \DsSolGptCTot & \DsSolGptCUNK & \DsSolGptCTP & \DsSolGptCTN &\DsSolGptCFP & \DsSolGptCFN &  \DsSolGptCAcc\% & \DsSolGptCPrec\% & \DsSolGptCRec\% & \DsSolGptCSpec\% & \DsSolGptCFOne\%  \\ \hline 
		
		\rowgptC	\datasetNoSolCMC & \gptC & \DsNoSolGptCTot & \DsNoSolGptCUNK & \DsNoSolGptCTP & \DsNoSolGptCTN &\DsNoSolGptCFP & \DsNoSolGptCFN &  \DsNoSolGptCAcc\% & \DsNoSolGptCPrec\% & \DsNoSolGptCRec\% & \DsNoSolGptCSpec\% & \DsNoSolGptCFOne\%  \\

		\rowgptQ	\datasetSolCMC & \gptQ & \DsSolGptQTot & \DsSolGptQUNK & \DsSolGptQTP & \DsSolGptQTN &\DsSolGptQFP & \DsSolGptQFN &  \DsSolGptQAcc\% & \DsSolGptQPrec\% & \DsSolGptQRec\% & \DsSolGptQSpec\% & \DsSolGptQFOne\%  \\ \hline 
		
		\rowgptQ	\datasetNoSolCMC & \gptQ & \DsNoSolGptQTot & \DsNoSolGptQUNK & \DsNoSolGptQTP & \DsNoSolGptQTN &\DsNoSolGptQFP & \DsNoSolGptQFN &  \DsNoSolGptQAcc\% & \DsNoSolGptQPrec\% & \DsNoSolGptQRec\% & \DsNoSolGptQSpec\% & \DsNoSolGptQFOne\%  \\
		\hline
		
	\end{tabular}
}
\end{table}

\Cref{table:resBySubDSCVL} shows the results of the analysis.
Under all  metrics,
\gptQ performs considerably worse than \solcmc and \certora on their respective datasets, while
\gptC performs considerably better than \certora on all metrics, 
and better than \solcmc on all metrics except specificity (and it wins only by a narrow margin on precision).
This is due to the low number of FPs in \solcmc.
Remarkably, \gptC's performance is consistent across the restricted sub-datasets and their complements. 
This indicates --- perhaps surprisingly --- that the properties not expressible in the two verification tools are not inherently harder for \gptC.

Besides correctly identifying the validity of a property, 
the value of a verification tool lies in the guarantees it can provide about the answer (\ie, a counterexample or proof certificate). 
%
In \Cref{sec:RQ2}, we analyzed the answers of \gptQ and \gptC.
Here, we discuss how they compare with those of \solcmc and \certora.
Both tools, when returning \True, do not provide any explanation:
the guarantees just lie on their correct implementation and on their symbolic interpretation of the Solidity semantics
(but, as observed before, some approximations they make are unsound).
When returning \False,  
\solcmc tries to return a sequence of transactions 
that leads the contract in a state where the invariant is violated. 
In most experiments, however, it did not produce any counterexample; 
moreover, it may return FNs in certain cases in which it abstracts from the Solidity semantics~\cite{solcmc-main}.
%
%
Certora, on the contrary, over-approximates the set of reachable states: its counterexamples 
are states that violate the property, but they are not guaranteed to be reachable, thereby producing spurious counterexamples~\cite{certora-expr}. 
This explains the high number of FNs in~\Cref{table:resBySubDSCVL}.



%
%
%
%
%
%
%
%
%
%
\section{RQ4: LLMs for verification auditing in the wild}
\label{sec:RQ4}

In order to assess whether \gptC can be useful in real-world auditing scenarios, we selected the two most recent formal verification auditing reports made available by \certora at the time of writing: Safe v1.5.0 and infiniFi Protocol~\cite{CertoraReports}.
The Safe contract is a smart account wallet~\cite{githubSafe}, 
while infiniFi is a protocol that offers yield-bearing stablecoin deposits~\cite{githubInfinifi}.

The reports contain properties written in natural language,  each of which is associated to a formal specification written in CVL. 
%
%
%
%
We feed \gptC the properties written in natural language, and ask it to verify them with the same prompt used in the other experiments (\ie, the one 
in \Cref{app:prompt}).
%
%
%
We stress that, since \gptC knowledge cut-off is September 30th 2024~\cite{gpt5cutoff}, we can 
\change{assume} that the model \change{does not use} 
information found in the reports. 
This is further validated by the fact that \gptC  often disagrees with \certora. 

In general, 
\change{determining the ground truth of the properties is extremely challenging,}
as these contracts consist of thousands of lines of code, besides those in external dependencies.
Hence, reliably establishing the validity of 
\change{all the}
properties is rather unfeasible, forcing us to recur to a more empirical approach.

We carefully analysed \gptC answers, with particular attention to properties
over which \gptC disagrees with \certora.
%
%
First, we examined the CVL specification, to verify whether it is consistent with the property written in natural language; 
if not, we rephrased the property accordingly, and re-ran \gptC over the newly formulated property.
If the cause of disagreement is not an inconsistency between the formal specification and the property description, 
we then analysed the counterexample provided by one of the two tools.
%
%
%
%
In particular, since the states that \certora provides as counterexamples lie on an over-approximation of the set of reachable states, we had to ensure that the returned state is indeed reachable,
in order to confirm the violation. 
In order to do that, we paired manual inspection with new additional queries to \gptC, feeding it the \certora's state and asking it to provide a trace that starts from the deployment of the contract and leads to the state under analysis. 
\change{For a sample of the properties deemed as \False by \gptC but \True by \certora,  we produced Proofs of Concept (available in \cite{anonymous_repo}) written in Forge~\cite{Forge2025} that validate 
\gptC's answers.}




%
We summarise below the results of our experiments (available in~\cite{anonymous_repo}), and then discuss our findings.
An extensive report of the results is in~\Cref{sec:App:RQ4}.

\paragraph{Summary.}
\label{sec:RQ4Summary}
%
%
%


The Safe report contains 25 properties, 
all with reported \True status according to \certora, 
while 
the Infinifi report contains 47 properties, 8 of which are reported to have \False status and 41  \True status, according to \certora.

We summarize our observations as follows:
\begin{itemize}
	\item \emph{Property/Spec mismatch:} 
	In 12 cases, \gptC and \certora disagree due to a mismatch between the property description and the CVL spec, 
	either due to  assumptions made in the spec but missing in the property,
	or due to typos (\eg  ``$\leq$'' in the spec but ``$<$'' in the property).
	In all these cases, manual inspection confirms that \gptC results are correct \wrt the property description.
	If we rectify the  properties  to match the CVL spec, \gptC returns results consistent with \certora's. 
	
	\item \emph{Munged contract:}
	 In  6 cases, \gptC and \certora disagree due to the fact that  \certora  was run on a munged version of the contract, in which delegate calls were disabled.
	 This 
	 was not specified in the report,
	 and we had to inspect the CVL run to notice it. 
	 Manual inspection confirms that the violations found by \gptC indeed hold on the original contract. 
	 If we instruct \gptC to disregard delegate calls, it returns results consistent   with \certora's. 
	\item \emph{Cryptographic assumptions:} In 1 case, \gptC returns \Unknown as the validity of the property depends on a cryptographic assumption, 
	which the Certora Prover makes internally~\cite{certora-hasing}.
	If we  add this assumption to the property, \gptC returns a  result  consistent with \certora's.
	\item \emph{Math errors:} In 4 cases, \gptC makes errors involving numbers in scientific notation. 
	Instructing it to always convert to decimal notation before performing computations was sufficient to prevent it from repeating the errors.
	\item \emph{Missing violations:} In 3 cases, \gptC does not spot a violation. In 1 case, it does not detect unstable rounding behaviors; in 2 cases, it overlooks hooks calls that can cause the violation (we note that, if we instruct \gptC to ``\verb|Pay attention to hooks|'', it can spot the violations).
	
	\item \emph{Validating Certora's cex:} For each property claimed as \False by Certora, we fed the returned state to \gptC and asked it  to verify whether the state is reachable or not, and, if yes, to provide a trace that starts from deployment. 
	In all cases except one, \gptC confirms that the state returned by Certora is indeed reachable, validating the violations found.
	In the remaining case, even manual inspection was not able to verify whether the state is indeed reachable; in this case, the correctness of the violation found by Certora is questionable. 
	We refer to \Cref{sec:App:RQ4Inf} for a detailed discussion.
	
\end{itemize}

\paragraph{Discussion.}

\label{sec:RQ4Disc}

We first note that, in most cases, \gptC returned answers that match what we believe to be the respective ground truths.
In particular, in most of the cases in which a property is violated, \gptC was able  to detect the violation and  return a valid counterexample.
Furthermore, we observed a powerful synergy between \gptC and Certora for the validation of  violation counterexamples.
Indeed, Certora returns states that are not guaranteed to be reachable (and the tool itself is not able to prove reachability, in general).
\gptC, on the other hand, is quite good at producing detailed traces that start from contract deployment to reach a given state.
	
Another positive role \gptC excels at is spotting inconsistencies between the  description of properties in natural language and the formal specifications.
This is a rather important aspect of the auditing process, 
as the quality of a report also depends on the correct explanations of what has been verified, precisely (readers are not supposed to investigate formal specifications themselves, but rather trust the property descriptions available in the report).
	
 A weak point of \gptC is that it struggles with scientific notation, which causes it to fail to find vulnerabilities related to subtle mathematical dynamics.
 
We also observed that \gptC sometimes improves its answers when given targeted hints, such as paying particular attention to some features (\eg hooks) or to some methods.
This is not very different from how auditors use Certora, which often needs to be instructed to only analyse certain parts of the code, or to adopt simplifying assumptions without to avoid timeout, \etc
This further validates that human guidance still plays a relevant role in the auditing process, regardless of the specific tool used.







\section{Related work}
\label{sec:related}

Research on smart contracts analysis has mainly followed two directions: vulnerability detection and formal verification.
The former focusses on finding bugs in contract code by leveraging recurring vulnerability patterns.
In this setting, there exist large datasets of contracts tagged with their vulnerabilities that are used as benchmarks to evaluate new analysis techniques~\cite{smartbugs_github,defihacklabs2024,wang2024attackdb}.
Formal verification instead focusses on proving that a contract satisfies  specific properties of its business logic (or showing counterexamples otherwise). 
This approach is pursued through different perspectives, often related to the class of expressible properties.
For example, VeriSolid~\cite{Nelaturu23tdsc} focusses on CTL properties; 
SmartPulse~\cite{Stephens21sp} on past-LTL properties;  
Solvent~\cite{Solvent} on strategic properties of the form ``there exists some transaction leading to a certain state''.
Compared to \solcmc and \certora, however, these tools are not yet directly usable in the wild, as they typically support restricted fragments of Solidity.
%
Unlike the vulnerability detection setting, no large datasets exist for benchmarking formal verification tools.
The largest available dataset so far contains only 323 verification tasks~\cite{BFMPS24fmbc}; whereas the dataset developed for this work comprises \Nvertasks\xspace tasks.

LLMs offer a complementary path to contract analysis: thanks to their understanding of natural language and 
pattern-driven problem-solving approach,
they may help overcoming some of the limitations of formal verification tools.
%
%
\change{A1~\cite{Gervais25ai} and ReX~\cite{Prompt2Pwn} leverage LLMs
--- paired with testing and concrete execution ---
to generate exploit  against on-chain contracts, showing that they are are able to reproduce exploits for several historical vulnerabilities. 
}
%
The role of LLMs
in the context of smart contract auditing has been first analysed \cite{StillNeedManualAudit}, where \gptQ and Claude are queried to detect the presence of a pre-defined vulnerability patterns (\eg reentrancy, delegatecall injection, \etc). 
%
%
%
SmartGuard~\cite{SmartGuard} addresses vulnerability detection by combining the retrieval of semantically similar code with chain-of-thought generation and a self-check architecture, 
validating their technique on the SolidiFI benchmark~\cite{SolidiFI}. 
%
The work~\cite{Liu2025BCRA} fine-tunes ChatGPT on 7 logical vulnerabilities (\eg, price oracle manipulation, privilege escalation), and experiments on a dataset of audit reports.
LLM-SmartAudit~\cite{LLM-SmartAudit} employs a multi-agent conversational framework with a buffer-of-thought mechanism, 
addressing 10 types of common vulnerabilities (reentrancy, overflow, \etc).


Compared to the previous works, ours takes a different perspective: rather than focussing on detecting known vulnerability types\change{, and possibly use them to synthesize exploits}, we investigate whether LLMs are capable of reasoning about arbitrary, contract-specific properties,
and whether they can provide convincing explanations for their answers.
 

During the construction of our dataset, we observed that crafting properties that precisely capture the desirable contract behaviors is quite challenging. To address this issue, some works have proposed LLMs as an aid to automate the generation of such properties.
PropertyGPT~\cite{liu2025propertygpt} uses LLM-based in-context learning to generate properties from contract code, targeting in a custom specification language similar to \certora's CVL. 
The framework DbC-GPT~\cite{leite2024extracting} proposes to use LLMs to generate post-conditions in the language of solc-verify~\cite{Hajdu19vstte}.
%
A limitation of both works is that they cannot deal with arbitrary properties, since after generation they still resort to symbolic tools for verification.
As noted before, this rules out relevant classes of properties (\eg, strategic ones), which could provide more precise insights on the desirable contract dynamics.
Indeed, it has been observed that ``global'' properties (such as metamorphic or strategic properties) usually 
yield a higher return of investment than ``local'' ones, such as function specs~\cite{Xu24langsec}.




While our work only hints at the potential of combining LLMs with formal methods, other studies --- in domains outside of smart contracts --- have already shown that such integration can be quite effective in practice.
%
For example, ESBMC-AI~\cite{tihanyi2025new} integrates LLMs with bounded model checking to detect and repair common flaws.
The work~\cite{kamath2024leveraging} shows that LLMs can aid in formal verification tasks, such as inferring loop invariants and proving safety properties.
%
\section{Limitations}
\label{sec:limitations}

\paragraph{Soundness.}

The use of LLMs in the verification context cannot replace --- nor it should be intended as a substitute for ---  traditional formal verification approaches.
Indeed, LLMs' predictions and explanations --- although statistically accurate --- are not supported by sound reasoning and are not certifiable.
Nevertheless, both the experiments on the synthetic dataset and those on real-world reports suggest that LLMs can be profitably used in combination with symbolic techniques, \eg concrete PoCs that refine and certify the LLM-produced counterexamples.  

\paragraph{Limitations of the dataset.}

Although our dataset includes a wide range of property classes --- including  state invariants, single- and multi-transition invariants, function specs, metamorphic and strategic properties --- it does not encompass all relevant classes of properties. 
In particular, some contracts, such as gambling games and certain DeFi protocols, rely on incentive mechanisms whose correctness can only be analysed in a game-theoretic setting. 
For example, in an $N$-players lottery, one would be interested in proving \emph{fairness} properties, such as that each honest player will have a non-negative payoff~\cite{Andrychowicz16cacm,Miller17eurosp,Chatterjee18esop}.
%
Additionally, DeFi protocols are usually designed to achieve complex economic properties (\eg, ``market efficiency'' for AMMs). 
These kinds of properties are beyond the scope of existing industrial verification tools such as SolCMC and Certora. 
Investigating whether LLMs can provide reliable predictions and explanations for these classes of properties is a largely unexplored direction for future work.

Additional limitations of our dataset concern its scale, the subset of Solidity features it explores, and its coverage of real-world contract patterns.
In particular, the contract variants included in the dataset do not expose some potentially problematic constructs such as the gas mechanism, loops, and delegate calls. 
Moreover, the included contracts do not cover some patterns used in practice, such as proxies, factories, and cross-chain interactions~\cite{Zhang24raid}.  
Assessing the LLMs' understanding of these features is another direction for future work.

\section{Conclusions}
\label{sec:conclusions}

In this paper, we have investigated the use of LLMs for smart contract verification.
Our systematic evaluation over a large dataset of verification tasks, 
as well as  experiments on two real-world smart contracts audits,
substantiate that 
state-of-the-art reasoning-oriented LLMs such as \gptC can be surprisingly effective at the task.
Due to their immediacy and ease of use, this can significantly push forward  the accessibility of verification to a wider audience,
enhancing smart contracts security.
%
%
%
%
We also observed great complementarity with formal verification tools, 
suggesting hybrid approaches as the next frontier. 

While this work paved the way to the use of LLMs for smart contract verification, several research questions remain open. 
A key challenge is how to improve the certifiability of LLMs answers, 
thereby reducing FPs and FNs.  

For \False-answers, our current approach is to ask the LLM to generate  natural-language counterexamples that illustrate the violation. While this provides auditors with useful intuitions about why a property is violated, it does not guarantee that the example is reproducible on the actual blockchain. To reduce such false negatives,
we could instruct the LLM to provide a concrete PoC, \eg in the form of an executable Hardhat script (which we leveraged for the manually-crafted PoCs in our dataset).

For \True-answers, ideally we would like the LLM to produce a machine-checked proof that the property holds in some proof assistant (\eg, in Rocq, Lean, Isabelle).
While we see this as a challenging long-term goal, there are already a few works that may help in reducing the gap towards it.
The work~\cite{Marmsoler25fac} formalises an executable semantics of a fragment of Solidity in the Isabelle/HOL proof assistant.
Leveraging such semantics, it is possible to craft machine-checked proofs for \True-properties.
While this is done manually in~\cite{Marmsoler25fac}, the long-term goal would be to exploit the reasoning capabilities of recent LLMs to assist humans in writing such proofs --- if not even generate them autonomously (in other domains, there is already ongoing research~\cite{ProofAutomationLLMs,cao-etal-2025-informal}).

%






\iftoggle{anonymous}{}{\paragraph{Acknowledgments}

Work partially supported by project SERICS (PE00000014)
under the MUR National Recovery and Resilience Plan funded by the
European Union -- NextGenerationEU, by PRIN 2022 PNRR project DeLiCE (F53D23009130001) and by project
EnergyCertifier (F23C23000270008)
funded  by Aut.~Reg.~of Sardinia / Sardegna Ricerche on the funding call ``Aiuti per Progetti di Ricerca e Sviluppo -- Settore ICT'' (2022).
The authors thank Francesco Stori and Niccolò Viale for contributing to the construction of the dataset. }

\bibliographystyle{splncs04}
\bibliography{main}

\appendix
\section{Prompt}
\label{app:prompt}

The text below is the Python string that defines the prompt within the script used to call the GPT API. The prompt file, including the script and other resources used for the experiments, as well as the resulting outputs, is publicly available in the project repository~\cite{anonymous_repo}.

\begin{verbatim}
You are an expert in Solidity smart contracts and formal verification. 

You will be given: 
1. A Solidity smart contract. 
2. A property to verify on the contract, expressed in natural language. 

Your task: 
- Analyze the contract carefully.
- Determine whether the contract ALWAYS satisfies the given property.
- If the property is FALSE, you MUST provide a concrete counterexample, 
for instance showing a state and/or a sequence of transactions that 
violate the property.
- If you are NOT SURE, respond ""UNKNOWN"".

Think step by step internally about whether the property holds, 
but DO NOT include your reasoning in the output.

In your reasoning, consider the following technical assumptions:
- ignore gas costs. For example, if a user sends a transaction 
that is reverted, ignore the gas paid by the user;.
- when in a property we refer to a ""user"" of the contract, we implicitly
assume that the user is not the contract under analysis itself;
- there is not (and there will never be) enough ETH in circulation to 
cause an overflow to a uint variable storing the balance of an address.

Always respond ONLY in the following format:
ANSWER: [TRUE | FALSE | UNKNOWN]
EXPLANATION: <brief explanation citing relevant lines or functions>
COUNTEREXAMPLE: <if ANSWER=FALSE, provide one; otherwise write ""N/A"">
---
Smart Contract:
{code}
Property:
{property_desc}   
\end{verbatim}

\section{RQ4: LLMs for verification auditing in the wild}
\label{sec:App:RQ4}

We  detail the results of our experiments, respectively for the Safe contract (\Cref{sec:App:RQ4Safe}) and the InfiniFi protocol (\Cref{sec:App:RQ4Inf}).
The complete report of all the experiments performed  can be found in the public repository~\cite{anonymous_repo}.


\subsection{Safe}
\label{sec:App:RQ4Safe}
The report contains 25 properties, 
all with reported \True status
according to Certora.

We carefully analyzed the properties for which \gptC returned an answer different from that returned by Certora. 
Such set contains 14 properties overall: 13 reported as \False  and 1 reported as \Unknown.
We investigated the causes of disagreement.
We organize our findings as follows:
\begin{itemize}
	\item \emph{Property/Spec mismatch:}  In 7 cases, the property written in natural language did not match the formal specification written in CVL. 
	\begin{itemize}
		\item in 4 cases, the CVL specification assumes that the mentioned call does not revert;
		without such assumption, the properties would  indeed be violated.
		\item in 1 case, the CVL specification assumes the value of a given parameter;
		without such assumption, the property would  indeed  be violated.
		\item in 1 case (\propname{setSafeMethodChanges}), the property description is ambiguous: it mentions that the call to 
		a method
		 changes the value of a variable; however, what is guaranteed is only that the call to the method sets the variable to the new specified value, which can be equal to the previous value.
		 The report contains another property which  precisely expresses this condition (\propname{setSafeMethodChanges}).
		The CVL specifications of the two properties are indeed exactly the same. 
 \item in 1 case (``\emph{handlerCallableIfSet}''), the CVL specification assumes that a given handle is a dummy method. 
		 Furthermore, we need to make the assumption that the choosen handler selector does not collide with that of any other  existing function. 
	\end{itemize}
	
	\item \emph{Munged contract:} \gptC has found  6 properties which are violated due to the possibility of executing delegate calls.  
	We inspected the counter-examples found, and they are indeed correct.
	The reason why the Certora Prover states that these properties are satisfied is because the prover was actually run on a munged version of the contract, in which delegate calls are 
	not allowed.
	This assumption was nowhere specified in the report. 
	
	
	\item \emph{Cryptographic assumptions:} In 1 case, \gptC returned \Unknown, since the property under analysis only holds if we assume that keccak256 values cannot collide;
	the Certora Prover internally makes this assumption~\cite{certora-hasing}.
\end{itemize}

For all these properties, we have introduced new corresponding  properties,
adding the assumptions previously not  specified. 
On all these new properties, \gptC returned \True, matching Certora's outputs.
Note that, in all these cases, we  never changed the code of the contract, but only the property description.

\subsection{InfiniFi}
\label{sec:App:RQ4Inf}

The report contains 47 properties, 8 of which are reported to have \False status and 41  \True status, according to Certora.

We analyzed the set of properties over which \gptC returned an answer different from that returned by Certora (which contains 9 properties), 
and the set of properties over which both Certora and \gptC returned \False (which contains 3 properties)

We organize our findings as follows:
\begin{itemize}
	\item \emph{Certora \True\ / \gptC \False:} In 5 cases, Certora reported \True while \gptC reported \False:
	\begin{itemize}
		\item 
		in 4 of these cases, there is a mismatch between the property description and the CVL specification. 
		In particular: 
		\begin{itemize}
			
		\item in 1 case the property description uses a strict inequality ($>$), while the CVL spec uses a non-strict inequality ($\geq$)
		\item in 1 case the property description excludes reverting, while the CVL spec excludes both reverting and returning $0$,
		\item in 2 cases the CVL spec excludes calls to methods that are not excluded in the property description
		\end{itemize}
	In all these cases, the property as written in natural language is indeed violated, and the counterexamples returned by \gptC are correct.
	If we re-run \gptC after adding the respective assumptions to the property descriptions, it returns \True on all cases.
		\item in 1 case, \gptC commits a math error involving scientific notation (\eg it states that $\code{6e18 / 3e18 = 2e18}$). 
		We instruct \gptC how to deal with numbers in scientific notation
		adding to the prompt ``\emph{Be careful: whenever you have to perform arithmetical computations concerning numbers in scientific notations (e.g. \code{1e10}, \code{6.3e14}, etc), you must FIRST convert them to decimal notation (e.g. \code{10000000000}, \code{630000000000000}), and only THEN perform arithmetical computations.}''
		After this instruction, \gptC returns again \False, however providing a correct counterexample.
		After inspection, we noted that the CVL specification makes an additional assumptions, \ie that USDC prices are around 12 orders of magnitude greater than iUSDC prices. 
		If we add this assupmtion to the property description, \gptC returns \True.
		
	\end{itemize}
	\item \emph{Certora \False\ / \gptC \True}: In 4 cases, Certora reports \False while \gptC returns \True:
	\begin{itemize}
		\item in 3 cases, after we asked \gptC to produce a complete trace that leads to the state provided by Certora as a counterexample, 
	 	\gptC returned a complete and correct trace, validating Certora's counterexample.
	 	In 2 of these 3 cases, the trace involves hooks, 
	 	which have probably been overlooked by \gptC on the first run.
	 	If we suggest \gptC to be careful about hooks 
	 	(\ie we add ``\emph{Pay attention to hooks.}'' to the prompt), 
	 	it returns \False, providing a correct counterexample.
	 	In the third case, the property is violated due to unstable rounding implementation: 
	 	\gptC limited math 
	 	reasoning
	 	is probably the reason why it was not able to spot the violation.
		\item in 1 case,  after we asked \gptC to produce a complete trace that leads to the state provided by Certora as a counterexample, 
		\gptC claimed that the state is not reachable. 
		We manually inspected that state,
		and have indeed some reservations regarding the reachability of the state. 
		The following explanations has necessarily to  enter into technical details.
		 The CVL spec, in the precondition checks the total reward not through \code{point.totalRewardWeight}, but through \code{totalRewardWeight()}, which is equal to \code{point.totalRewardWeight.mulWadDown(slashIndex)}, i.e. it returns a value already approximated.
		 As well, it checks the users weights not through \code{position.fromRewardWeight}, but through \code{rewardWeight()}, which is equal to \code{position.fromRewardWeight.mulWadDown(slashIndex)}, i.e. it returns a value already approximated.
		 Indeed, Certora returns a state in which \code{UnwindingModule.globalPoints[0x277c].totalRewardWeight=1.33e18}, which is exceeded by the \code{rewardWeight} of a single user, \\ {\code{UnwindingHarness.positions[0x274c].fromRewardWeight = 1.5e18}}.
		 However, the precondition is satisfied, since \code{totalRewardWeight() =} \\ \code{ rewardWeight(userA) = 4}.
		 Whether the state returned by Certora (i.e. where \code{totalRewardWeight} is lower than a user \code{fromRewardWeight}) is reachable, is unclear.
		 In the report, no elaboration is provided, besides an argument about rounding error, 
		 which however does not imply the state to be reachable,
		 but only that, if the state is reachable, then there exists a transaction that can violate the property. 
		 %
		 Our claim is that the desired  property to verify is actually that \code{point.totalRewardWeight} is greater than the sum of the \code{x.fromRewardWeight} for all users.
		
	\end{itemize}
	\item \emph{Certora \False\ / \gptC \False}: In 3 cases, both Certora  and \gptC return \False:
	\begin{itemize}
		\item in 1 case, \gptC finds a counterexample similar to the state returned by Certora
		\item in 2 cases, \gptC's counterexample is incorrect due to math errors involving scientific notation. 
		If we instruct \gptC to convert numbers from scientific notation to decimal, it returns \True. 
		However, if we ask it to produce a complete trace that leads to the state provided by Certora, 
		it returns a complete and correct trace that validates Certora's counterexample.
		The states returned by Certora involve large numbers that only differs by small digits.
		The reason why \gptC was not able to find the counterexamples straight ahead is probably due to its limited math capabilities involving large numbers.
	\end{itemize}
\end{itemize}





\end{document}